\documentclass[aps,prl, reprint,  onecolumn, amsmath, superscriptaddress]{revtex4}
\usepackage{graphicx}
\usepackage{amsfonts}
\usepackage{bm}
\usepackage{epstopdf}
\usepackage{ulem}
\usepackage{comment}
\usepackage{xcolor}
\usepackage{amsmath}

\newcommand {\be} {\begin{equation}}
\newcommand {\bea} {\begin{eqnarray} }
\newcommand {\bean} {\begin{eqnarray} \nonumber}
\newcommand {\ee} {\end{equation}}
\newcommand  {\eea} {\end{eqnarray}}
\newcommand {\nn} {\nonumber}
\newcommand {\ximu} {\xi^{(\mu)}}
\newcommand {\bximu} {\bar\xi^{(\mu)}}
\newcommand {\mH} {{\mathcal H}}
\newcommand {\mM} {{\mathcal M}}
\newcommand {\mS} {{\mathcal S}}
\newcommand {\mZ} {{\mathcal Z}}
\newcommand {\bs} {{\bm s}}
\newcommand {\ssa} {s^{(a)}}
\newcommand {\ssb} {s^{(b)}}
\newcommand  {\s} {\sigma}
\newcommand  {\la }{\lambda}
\newcommand  {\lan} {\langle}
\newcommand  {\ran} {\rangle}

\begin{document}
\title{Optical computation of a spin glass dynamics with tunable complexity}

\author{M. Leonetti}
\affiliation{Center for Life Nano Science@Sapienza, Istituto Italiano di Tecnologia, Viale Regina Elena, 291 00161 Rome, Italia.}
\affiliation{CNR NANOTEC-Institute of Nanotechnology, Soft and Living Matter Lab, P.le Aldo Moro 5, I-00185 Rome, Italy}
\author{E. H\"{o}rmann}
\affiliation{Department of Physics, University Sapienza, P.le Aldo Moro 5, I-00185 Roma, Italy}
\author{L. Leuzzi}
\affiliation{CNR NANOTEC-Institute of Nanotechnology, Soft and Living Matter Lab, P.le Aldo Moro 5, I-00185 Rome, Italy}
\affiliation{Department of Physics, University Sapienza, P.le Aldo Moro 5, I-00185 Roma, Italy}
\author{G. Parisi}
\affiliation{Department of Physics, University Sapienza, P.le Aldo Moro 5, I-00185 Roma, Italy}
\affiliation{CNR NANOTEC-Institute of Nanotechnology, Soft and Living Matter Lab, P.le Aldo Moro 5, I-00185 Rome, Italy}
\affiliation{INFN, Sezione di Roma I, P.le A. Moro 2, 00185 Roma, Italy}
\author{G. Ruocco}
\affiliation{Center for Life Nano Science@Sapienza, Istituto Italiano di Tecnologia, Viale Regina Elena, 291 00161 Rome, Italia.}
\affiliation{Department of Physics, University Sapienza, P.le Aldo Moro 5, I-00185 Roma, Italy}
\date{\today}
%

%
%
%

\begin{abstract}
Spin Glasses (SG) are paradigmatic models for physical, computer science, biological and social systems. The problem of studying the dynamics for SG models is NP-hard,  i.e., no algorithm solves it in polynomial time. Here we implement the optical simulation of a SG, exploiting the $N$ segments of a wavefront shaping device to play the role of the spin variables, combining the interference at downstream of a scattering material to implement the random couplings between the spins (the $J_{ij}$ matrix) and measuring the light intensity on a number $P$ of targets to retrieve the energy of the system.  By implementing a plain Metropolis algorithm, we are able to simulate the spin model dynamics, while the degree of complexity of the potential energy landscape and the region of phase diagram explored is user-defined acting on the ratio the $P/N=\alpha$. We study experimentally, numerically and analytically this peculiar system displaying  a paramagnetic, a ferromagnetic and a SG phase, and we demonstrate that the  transition temperature $T_g$  to the glassy phase from the paramagnetic phase grows with $\alpha$.
With respect to standard ``in-silico'' approach, in the optical SG interaction terms are realized simultaneously when the independent light rays interferes at the target screen, enabling inherently parallel measurements of the energy, rather than computations scaling with $N$ as in purely $\it{in-silico}$ simulations.
\end{abstract}



\maketitle

The solution of large combinatorial problems demands for novel hardware architectures enabling for faster and inherently parallel calculation.   An emerging trend is that of pairing an optical layer into a specific digital or analog computation scheme, in order to improve performance while reducing computational costs and processing times. Optical computing promises parallel processing and high bandwidth which may be eventually performed in free space, with limited power consumption (e.g., Fourier transform performed by a lens). Optical computation is an emerging scheme in quantum transport \cite{harris2017quantum}, quantum simulation  \cite{sparrow2018simulating}, and machine learning \cite{lin2018all}, and can be implemented on different platforms including free space,  photonic chips \cite{peruzzo2014variational} and optical fibers \cite{Roques-Carmes:19}. One of the advantages brought by optics is that certain operations can be  performed at the ``speed of light''. Indeed, the evaluation of a matrix product can be estimated in the time needed for a properly shaped light beam to pass through a diffractive pattern opportunely tailored to mimic the requested transfer matrix \cite{farhat1985optical}. By exploiting last generation optical modulation devices, millions of light rays can be driven simultaneously between several states and within a microsecond time frame, thus potentially providing a scalable optical platform that only need to be properly projected on to the relevant and computationally hard problem.

Spin glasses \cite{edwards1975} serve as prototype models, capable to provide nontrivial equilibrium and off-equilibrium phenomenology \cite{mezard1987spin,young1998spin}. In  particular, the dynamics in an energy landscape with many equilibrium states and the origin of (multiple) relaxation times in finite dimensional systems, are open questions in modern statistical mechanics \cite{fisher1988equilibrium,leuzzi2009ising,temesvari2010theising,janus2018aging}. Complex systems from
diverse  fields fall into the spin-glass universality class, like, e.g.,  brain functions \cite{amit1992modeling},  random lasers \cite{ghofraniha2015experimental,antenucci2016}, and quantum chromo-dynamics \cite{halasz1998phase}. Indeed,  novel methods for the calculation of the equilibrium states and of the dynamics of a spin glass system are highly desiderate.

\begin{figure}[h!]
\centering
\includegraphics[width=16 cm]{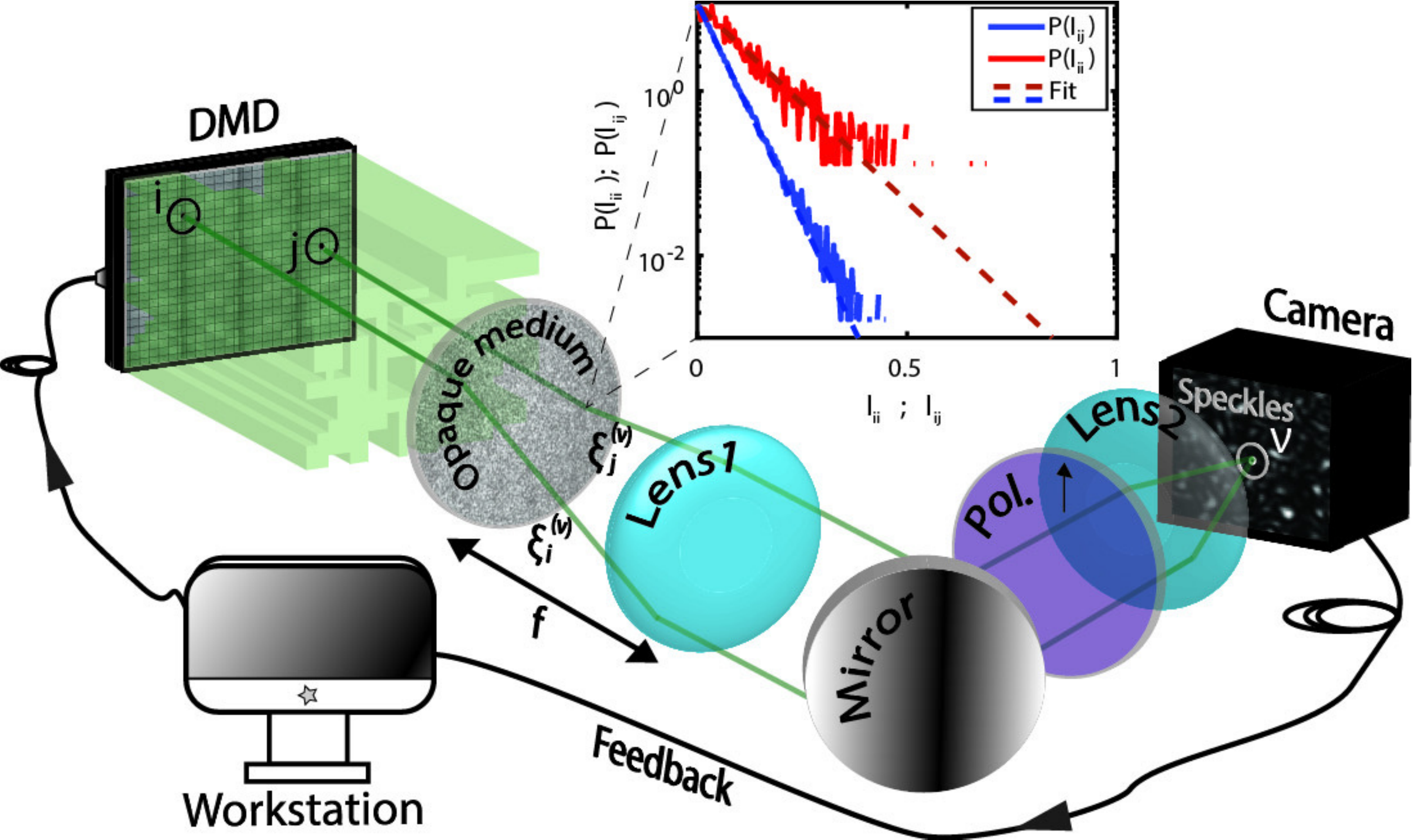}
\caption{A sketch of the experimental setup. Laser light (Azure Light 532 nm),  reflected by the de DMD, is then scattered by an opaque medium. The far facet of the scattering medium, is then imaged on a camera, after passing on a linear polarizer. The inset shows the measured $P(I_{\rm off~diag})$ and $P(I_{\rm diag})$. They have bene fitted with the function $P(I_*)=A_*\exp(-I_* / B_*)$. We retrieve $B_{\rm diag}=0.08$ and $B_{\rm off~diag}=0.04$. The values of the $B_*$ are consistent with the predicted behavior of  $I_{\rm diag}= v_{11}= |\xi_1|^2$ and $I_{\rm off~diag}= v_{12}+v_{21} = 2 v^R_{12}=2\Re[\xi_1\bar\xi_2]$ (see methods),
 that is  : $P(v^R_{ij}) = 1/(2 \sigma^2)  \exp^{ -|v^R_{ij}| / \sigma^2}$ and $P(v_{ii}) = 1/(2 \sigma^2)\exp^{ -v_{ii} /(2\sigma^2) }$.}\label{fig:setup}
\end{figure}

Here we propose an optical system able to compute the energy of a given spin-glass state. We integrated such optical layer onto a standard digital computation layer to realize an optical spin-glass (OSG) dynamics simulation. Our idea stems from the observation that the overall intensity $I=\sum_{\nu=1}^PI^{(\nu)}$ at $P$ given points $(\nu)$ on a screen placed at the downstream of a strongly scattering medium shone with  $N$ coherent light rays from a single laser, can be formally written as a spin glass Hamiltonian. Thus scattering, coupled with an adaptive optical element, has been proposed as an instrument to access the spin-glass dynamics and employed for low complexity (small $P$ values) simulations \cite{ErikHormannT}. Successively a similar approach has been employed to find ground states of large transmission matrices \cite{pierangeli2020scalable}.

As a starting  point for our experiment it is easy to observe that, in the simplified case in which the $i$th light ray field has a complex amplitude $a_i=A_ie^{\imath \phi_i}$, the single target contribution $I^{(\nu)}$ reads

\begin{eqnarray}
I^{(\nu)}&=&E^{(\nu)}E^{(\nu)\dag}=\left|\sum_{i=1}^N \xi^{\nu}_i a_i \right|^2
\label{eq:Inu}
=\sum_{i,j}^N \xi^{(\nu)}_{i} {\bar\xi}^{(\nu)}_{j} a_i \bar a_j
\\
\nonumber
&=&\sum_{i,j}^N    |\xi^{(\nu)}_{i}|  |\xi^{(\nu)}_{j}| A_iA_j  e^{\imath (\arg{\xi^{(\nu)}_{i}} -\arg{\xi^{(\nu)}_{j}} + \phi_i -\phi_j)}
\end{eqnarray}
where $\xi_i^{(\nu)}=|\xi^\nu_{i}|e^{\imath \arg{\xi^{(\nu)}_{i}}}$ are the complex transmission matrix elements from the $ith$ incoming beam to the target $\nu$.    Input illumination is controlled by a DMD with the superpixel method \textbf{(see methods)}. This approach enables to separate the input laser wavefront in many segments  (up to  $50^4$)  each one composed by 4  DMD mirrors. Each segment can be then programmed into one between two states each characterized by a phase factor $\phi_i=\pm \pi$, equivalent to an amplitude  factor of $S_i \in \{-1, 1\}$, so that the single target intensity (\ref{eq:Inu}) can be rewritten as
\begin{equation}
I^{(\nu)}=\sum_{ij}^{1,N} v^\nu_{i,j}S_iS_j \ ,
\label{I_SiSj}
\end{equation}
where all transmission matrix elements and input field amplitudes have been included in the coefficient
\begin{equation}
v^{(\nu)}_{ij}\equiv A_iA_j \xi^{(\nu)}_{i}{\bar \xi}^{(\nu)}_{j} \ .
\label{eq:v}
\end{equation}
Let us stress that though $v^{(\nu)}_{ij}$ are complex-valued, the intensity $I^{(\nu)}$ is always a real number because
$v^{(\nu)}_{ij}=\bar v^{(\nu)}_{ji}$ and the sum in Eq. (\ref{I_SiSj}) runs on all $i,j=1,\ldots,N$. Amplitudes $A_i$ are defined by the laser intensity. By using a laser with a  Gaussian beam and expanding it to retrieve an homogeneous distribution of the intensity on the active DMD  area, it is possible to approximate all the $A_i$, for any $i$,  to a constant over the DMD segments.
Maximizing the overall intensity with respect to DMD spin $\bm S$ configurations, finally corresponds to minimize the following Hamitonian
\begin{eqnarray}
\mathcal H[\bm S]&=& -\frac{1}{2} \sum_{ij}^{1,N} J_{ij}S_iS_j \ ,
\label{eq:H}
\end{eqnarray}
where we have introduced the interaction matrix
\begin{eqnarray}
J_{ij}&\equiv&\frac{1}{N}\sum_{\nu=1}^P v^{(\nu)}_{ij} \ ,
\label{eq:J}
\end{eqnarray}
properly rescaled with $N$ in order to provide thermodynamic convergence in the large $N$ limit also when the number of targets grows like the number of spins: $P=\alpha N$, with $\alpha=O(1)$.
Note that the matrix $J$ is a Hermitian matrix, $J_{ij}=J_{ji}^\dag$, and $\mathcal H[\bm S]$ is, therefore, real.

%
%
%

Equation \ref{eq:H} is, by all means, a spin glass Hamiltonian, more precisely it is  a generalization to complex  continuous valued patterns $\xi$ of the Hopfield model \cite{hopfield1982,hopfield1984}, for which the study of Amit, Gutfreund and Sompolinsky \cite{amit1985storing,amit1985spin,amit1987statistical}  predicts the existence of a low temperature/large $P$ spin-glass phase. That is, a phase with multiple degenerate frustrated thermodynamic  states, and critical slowing down dynamics approaching the glassy transition from the paramagnetic phase. To check this, we realized experimentally our optical simulation of a spin glass.
As shown in Fig. \ref{fig:setup} we exploited laser light form a stabilized Nd-YAG laser (\emph{Azure light 0.5 W}), to shine a Digital Micromirror Device (DMD, Vialux	V-7000) controlled by a CPU based workstation.

The same computer simultaneously monitors the Camera (\emph{Ximea CB013MG-LX-X8G3}  64 Gbit/s speed on the PCIe Gen3 lane) which collects light at the downhill of a scattering medium (Thorlabs ground glass diffuser, \emph{Thorlabs DG05-220}). A two lens system, enables to monitor in the camera plane the intensity corresponding to the output facet of the diffuser with a factor $8$ magnification (Focal of Lens1 is $25,4$ mm and focal of Lens2 is $200$ mm). To simulate  the system dynamics we implemented an opto/digital Monte Carlo Metropolis algorithm\cite{metropolis1953equation}, by flipping a random segment and accepting the change with a probability that depends on how intensity $I^{(\nu )}$ changes on the targets (a single target is a set of neighboring camera pixels) $\nu$. If $I^{(\nu )}$ increases, then the change is accepted, otherwise the change is accepted only with a  probability $p=e^{\Delta I/T}$ where $\Delta I$ is the measured variation in intensity following the spin/micromirror flip and $T$ is an user defined system temperature. The spin flipping is performed acting on the relative segment on the DMD (flipping time is 18 \textrm{${\mu}s$} ), while the spin coupling is realized optically thanks to the nearly instantaneous light propagation into the disordered medium, the intensity reading is performed through the camera, while the move acceptance is performed digitally by the computer.

\begin{figure}[b!]
\centering
\includegraphics[width=16 cm]{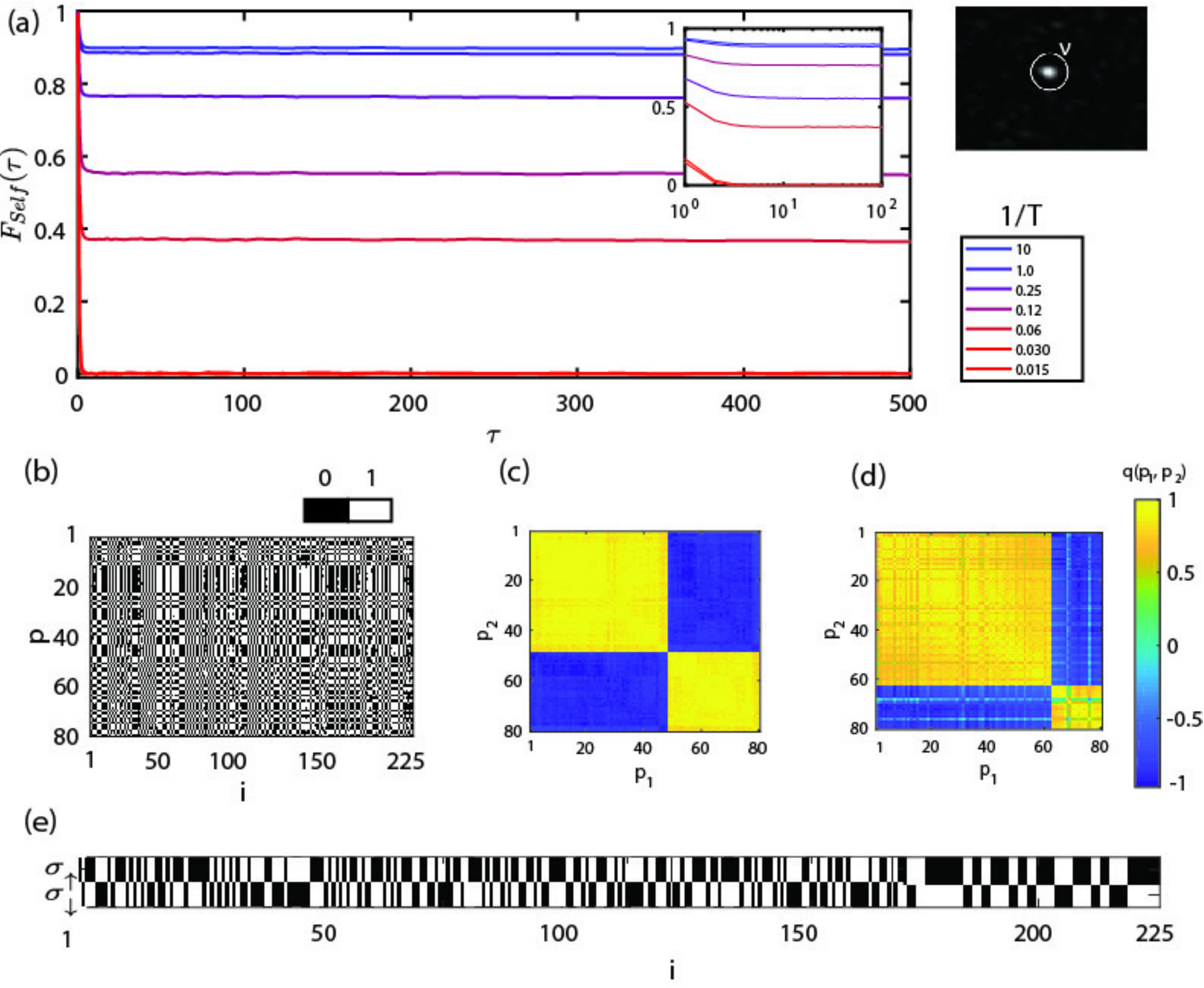}
\caption{ Experimental results for the $1$-target Regime. a)  $F_{self}$ as a function of  ${\tau}$, that is, the number of Metropolis  sweeps. The inset shows the first steps in logarithmic $\tau$  scale. b) Final states obtained after a series of repeated zero temperature simulations with the same disorder realization. The degree of complexity is low (just two opposite states are present). In c) and d) we report the overlap $q(p_1,p_2)$ -- defined in equation (\ref{eq:q}) -- between the configurations $\bm S^{(p_1)}$ and $\bm S^{(p_2)}$ of two replicated systems and sorted  by a $k$-means algorithm in order to be organized into two clusters. Replicas are labeled by the indexes $p_1$ and $p_2$ running on all the simulated systems.   The appearance of only two different yellow squares indicates that the configurations of the replicas thermalize into states that can be grouped in two clusters of very similar states one opposite to the other. d) is the same of c) for a different realization of disorder. In e) we report the two (opposite) states found in c). }
\label{fig:1Target}
\end{figure}

By storing spin configurations $\bm S(t)$ at each time step $t$  we are able to extract the temporal behavior of the connected autocorrelation function:
\begin{equation}
F_{\rm self}(\tau)=\frac{1}{N}\sum_i^N\langle S_i(t) S_i(t+\tau)\rangle_c;
\label{F_self}
\end{equation}
\noindent
where $\langle\ldots\rangle$  indicates the  averaging over Monte Carlo steps $t$.
As an instance, we show the case of an optical spin model with $N=225$ spins whose dynamics we simulated for $t_{max}=2\times 10^5$  Monte Carlo steps (each step consists of $N$ micromirror flips, with micromirrors selected uniformly and in random fashion) from which we extract the behavior of the correlation $F_{\rm self}(\tau)$ on a more limited temporal window for $\tau$ in order to perform a proper temporal averaging.

The results for a single target (number of targets $P=1$) is reported in Fig. \ref{fig:1Target}a, displaying $F_{\rm self}(\tau)$.
In this case, the correlation function decays rapidly to zero at high temperature and lowering the temperature its behavior crosses over to a power-law decay. This slowing down of the dynamics is called critical relaxation\cite{crisanti2007equilibrium,goetze2009complex,caltagirone2012critical,caltagirone2012criticalslo,caltagirone2012twostep}. Eventually, at  low enough $T$, $F_{\rm self}(\tau)$  tends towards a non-zero plateau. To check further the nature of the energy landscape, we run $80$ ($ p_1=1, \ldots, 80$ ) very short simulations ($\tau_{\rm max}\simeq 30$) with fixed target $\nu$ at zero $T$. In Fig. \ref{fig:1Target}b, we report the final states $\bm S_{\rm final}$ (the DMD square matrix has been reshaped to a 1D array and visualized as black or white squares respectively representing $S=-1$ and $S=1$) for each one of the simulations. We notice that, apart for a small fluctuation due to measurement noise on a limited set of  $i$'s, only two configurations $\bm S_{\rm final}$ are appearing. One, $\uparrow{\bm S}$, with an overall positive magnetization $1/N\sum_iS_i$,   and one, $\downarrow{\bm S}$, with negative magnetization,  as graphically reported reported in Fig. \ref{fig:1Target}e ). They are strongly anti correlated as shown by the $q$ parameter reported in Fig. \ref{fig:1Target}c and \ref{fig:1Target}d, for all the ${\bm S}_{\rm final}(p)$ obtained from $80$ replicas of the dynamics at $T=0$.  The parameter $q(p_1,p_2)$ shown there, represents the degree of similarity between the $\bm S_{\rm final}$ of two replicas of the system and is calculated as the normalized scalar product: $q(p_1,p_2)=\bm S_{\rm final}(p_1) \cdot \bm S_{final}(p_2)/N   $.  The  $q(p_1,p_2)$  matrix, appearing symmetric and with ones on the diagonal, has been visualized with  $p_1$ and $p_2$ sorted in order to clusterize similar $S_{final}$. In this favorable configuration,  states area appearing as yellow squares on the $q$ matrix. The two squares appearing in Fig. \ref{fig:1Target}c-d are states composed two the spin-reversed states confirming that in the case $P=1$ we have a transition to a ferromagnetic phase.

Indeed, the occurrence of only a pair of opposite states may be explained if we go back to the formal description of the $J_{ij}$ in Eqs. \ref{eq:v},  \ref{eq:H}, \ref{eq:J}, with $P=\nu=1$ target:
\begin{equation}
\mathcal H[\bm S]=-\frac{1}{2N}   \left| \sum_{i=1}^N   \xi_i^{(1)} S_i \right|^2 = -\frac{|\bm \xi^{(1)}\cdot \bm S|^2}{2N}
\label{I_simplified}
\end{equation}
in which we exploited the approximation of constant amplitudes for all the incoming light rays and we rescaled $A_i  \xi_i^{(\nu)} \simeq A  \xi_i^{(\nu)}  \to \xi_i^{(\nu)}$.
The Hamiltonian is just proportional to the scalar product of the spin configuration array and the array of the transmission matrix elements $\xi_i^{(1)}$ from the micromirror-created spins ($i$) to the target ($1$).
This is maximum for two configurations: the one maximizing the field along the transmission vector ($\uparrow$, see Fig. \ref{fig:setup}) and the other maximizing the field along the opposite direction. It is in fact well known that, given a certain transmission matrix, it is possible to find the best input configuration producing maximum intensity \cite{popoff2010measuring} and that this configuration is ``unique''.

\begin{figure}[h!]
\centering
\includegraphics[width=16 cm]{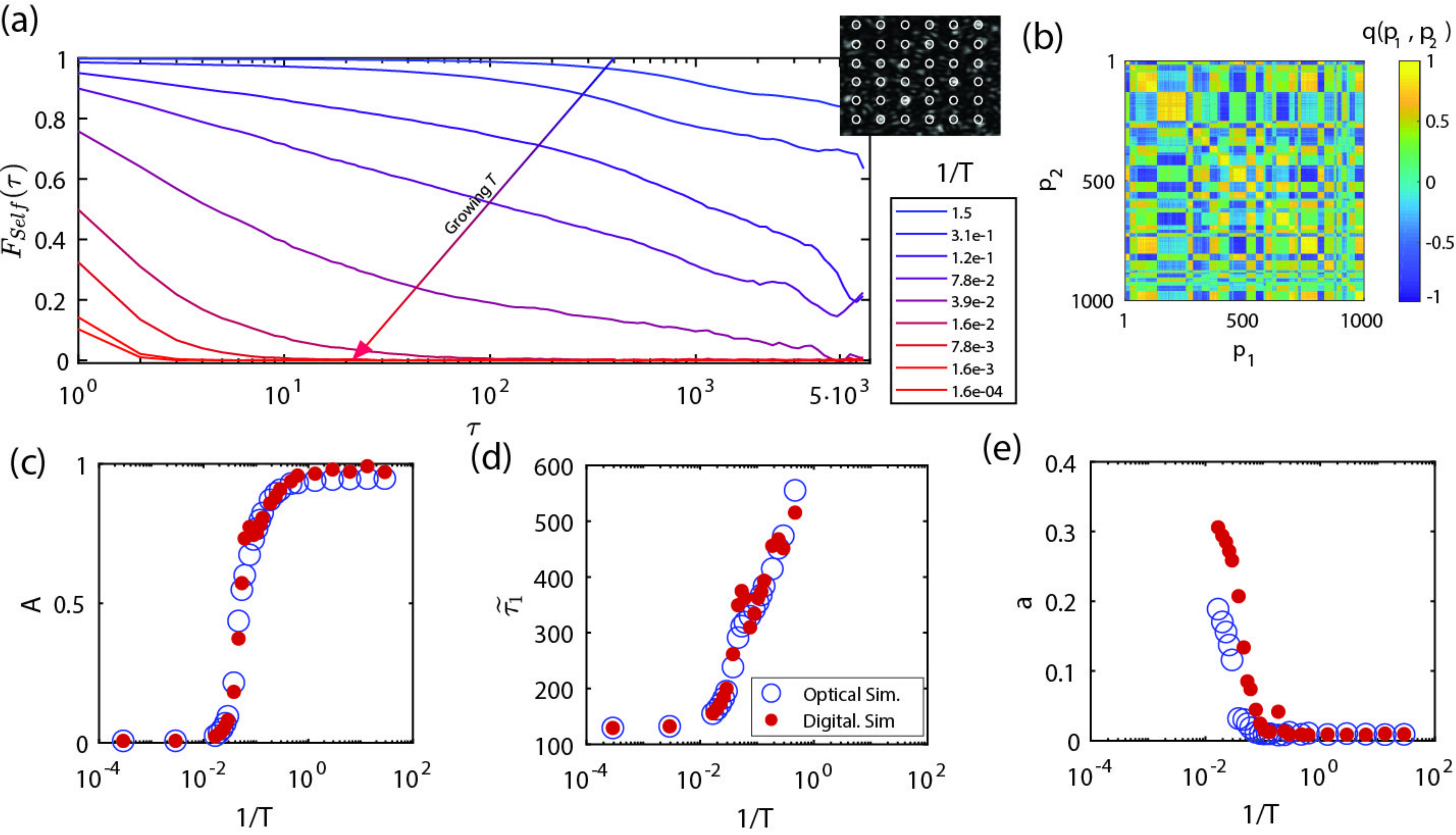}
\caption{  Results for $P= 36$, $N=225$.  In a) the  $F_{self}$ as a function of  ${\tau}$ indicating the sweeps.  In b) we report the degree of similarity obtained for clusterized states resulting  many optimizations at zero temperature. The phase space results clusterized in many states. In c) d) and e) we report respectively $A$, $\tau_1$  and $a$ as a function fo 1/T, comparing the results form the digital simulation (red full dots) with that of the optical simulation (open blue symbols) \textbf{(see methods)}. In panel d) $\tau_1$ is reported only in the region where the exponential contribution results a reasonable interpolation while in panel e) $a$ is reported only in the critical region. } \label{fig: 36tg}
\end{figure}

The interaction matrix $J_{ij} = \xi^{(1)}_{i} {\bar\xi}^{(1)}_{j}/N$
is a diadic matrix constructed on a single vector  $\bm \xi^{(1)}$ with the Hebb rule \cite{hebb1962organization}.
Resulting from a diadic matrix, our scattering system is, thus,  behaving as a sort of trained optical memory in which the pattern $\bm\xi^{(1)}$ is the memory learned by the network.
 In this neural network contest ``learning" means varying  the couplings.  The network is designed in such a way that the learnt patterns $\bm \xi$ are retrieved as  stable configurations, that is, in this example, the spin configurations $\uparrow{\bm S}$  and $\downarrow{\bm S}$.
In optical words, instead, when $P=1$ the experiment is identical to a standard wavefront shaping experiment \cite{vellekoop2007focusing} in which there exist a single spin configuration maximizing intensity \cite{PhysRevLett.104.100601}. The case where the $\xi$ are Boolean, rather than complex and Gaussian distributed,  is also called the Mattis model \cite{mattis1976}.

If $P$ is larger and, in particular, if it is so large to scale with the number of variables $N$, the situation is similar to the Hopfield model of Amit, Gutfreund and Sompolinksy  \cite{amit1985storing}. In that seminal paper the authors demonstrate that it is possible to store states into a neural network memory by exploiting a sum of diadic matrices generated from that many different vectors. However, there is a limit to the number of independent states that can be stored in such a kind of  memory\cite{buhmann1987influence}
 When this limit is surpassed, the multiple states start to combine, thus generating additional states to those related to the transmission patterns. All these states are minima of a complex energy landscape displaying a spin-glass phase in a given region of the phase diagram.
To experimentally engineer the complexity of the spin-glass phase in our optical system, then, the only thing that we need to do is to increment the number of targets. We, therefore, perform  optical Montecarlo simulation with the  Hamiltonian (\ref{eq:H}) and increasing $P$.

We perform a OSG simulations with $N=225$ and $P=4,9,16,25,36,225$ targets, corresponding to $\alpha\simeq 0.018$, $0.04$, $ 0.071$, $0.11$, $ 0.16$, $1$ (where  $\alpha$ is the ratio between the number of targets $P$ and the number  of degrees of freedom $N$). We report the results for $P=36$ in Fig. \ref{fig: 36tg}. In panel \ref{fig: 36tg}a we show the autocorrelation functions of the $\sigma$'s as a function of $\tau$. It is readily noticed that the $F_{\rm self}$ moves from an exponential to  a non-exponential relaxation in time as temperature decreases, that is a typical aspect of many complex systems \cite{kob1997dynamical,goetze2009complex,caltagirone2012critical}. In particular the $F_{\rm self}$ of the OSG with many targets can be typically described by the generic interpolating function approaching the critical point from high temperature:

\begin{figure}[h!]
\centering
\includegraphics[width=16 cm]{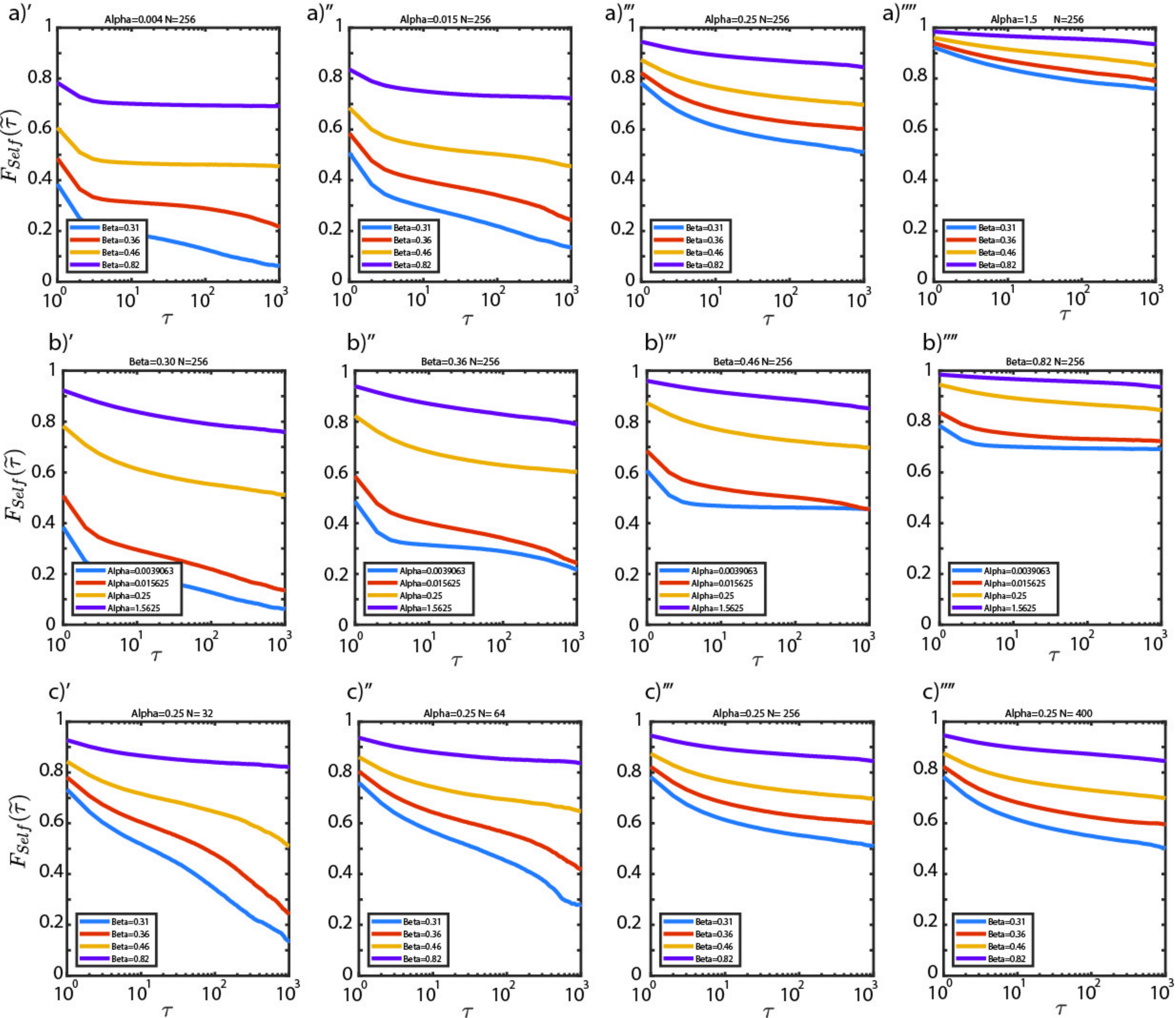}
\caption{$F_{Self}(\tau)$ for various values of $N$, $P$ and $\alpha$. From a${'}$ to a${''''}$:  each panel has a value of $\alpha$  (the same ) four temperatures. Panels from b${'}$ to b$''''$ report  the same data organized with a single temperature and four values of $\alpha$ per panel. Panels from c$'$ to c$''''$ instead report $\alpha$ fixed while $N$ (number of spins) is varied.} \label{Alpha}
\end{figure}

\begin{equation}
\nonumber
F_{\rm self}(\tau)=(1-A)e^{-\frac{\tau}{\tau_1}} + A \frac{C}{t^a} 
\end{equation}
\noindent
in which the first term represent the exponential relaxation time while the second term the power law decay appearing approaching the glassy phase, i. e., crossing over for large $\tau$. The $A$ parameter represents the relative weight of the two terms: the larger it is, the more the power-law-like is relevant. The behavior of the fitting parameters is reported in Fig. \ref{fig: 36tg}, estimated both by means of the  data obtained in standard ``digital" numerical simulations and with the optical Monte Carlo methods introduced in the present work. As expected, as temperature is decreased from the paramagnetic phase, we observe (i)  an increase of the  correlation time $\tau_1$, (Fig.  \ref{fig: 36tg}d), (ii)  a very sharp increase of the weight $A$ of the power-law decay contribution
in the critical region, (Fig.  \ref{fig: 36tg}c),  and (iii) a lowering of the power-law decay exponent $a$ in the critical region, (Fig. \ref{fig: 36tg}e). 

The critical slowing down of the relaxation is a signature of a multi-state (free-)energy landscape. Further evidence of complexity is also shown in an experiment in which multiple short relaxation dynamics at $T=0$ have been retrieved. In Fig.\ref{fig: 36tg}b) we report the parameter  $q(p_1,p_2)$ for a thousands  different simulations. Here the pattern shows  many different clusters  indicating that many equilibrium states are populating the potential energy landscape. We have been using $k$-means to organize data into $36$ clusters.

\begin{figure}[h!]
\centering
\includegraphics[width=12 cm]{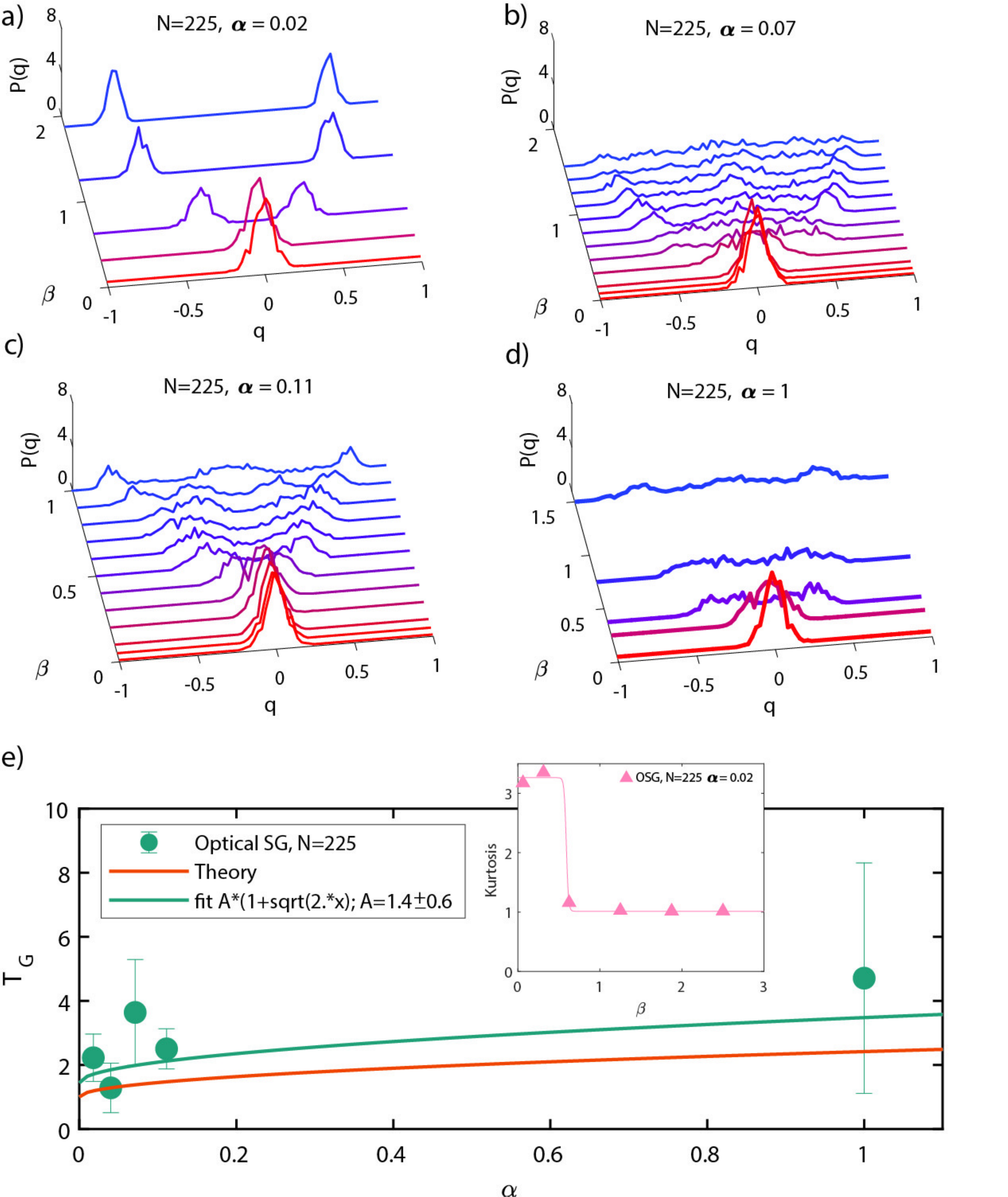}
\caption{Probability distribution of the Parisi parameter $q$ as a function of $\beta$, for N=225. Optical simulations have been performed for few values of $P$. a):$P=4$; $\alpha$=0.02    b):$P=16$; $\alpha$=0.07 c):$P=25;$  $\alpha$=0.11. $P(q)$ in a given $\beta$ , $\alpha$ configuration, has been obtained by performing 50 fast (25 sweeps) simulations (see methods). In panel d) we report simulations retrieved by performing 2 very long (8000 sweeps) simulations (see methods) for $P=225;$  $\alpha$=1.  In panel e) we report the estimated $T_G$ by means of optical simulations, together with the curve  $T_G=A(1+\sqrt{2\alpha})$ predicted from Replica theory for $N\to \infty$ (eq. \ref{Phasetranstion}). Optical simulation provides a value of A of $1.4 \pm 0.6$ compatible with theory (in which $A=1$). The inset of panel e) shows the Kurtosis of the $P(q)$ as a function of $\beta$. Continuous line is to guide the eye.  }\label{OVERLAP}
\end{figure}




In Fig.   \ref{Alpha} we compare the   $F_{\rm self}(\tau)$ for various values of $N$, the ratio $\alpha=P/N$  and the inverse temperature $\beta=1/T$. In the first set of panels, from a$'$ to a$''''$, we report the $F_{\rm self}(\tau)$ for four values of $\beta$ at fixed $N=256$ ( a different value of $\alpha$  is shown in for each panel). It is possible to note that at high $\alpha$, the decorrelation at any given temperature is strongly hindered, if not avoided at all, at the time-scales considered.


Panels from b$'$ to b$''''$ report  the same data (fixed $N=225$) organized at a single temperature in each panel and four values of $\alpha$ per panel. In panels from c$'$ to c$''''$, instead, we show the correlation function at fixed  $\alpha$ and varying  $N$ in each panel, demonstrating  that the behavior of  $F_{\rm self}(\tau)$, for the time window considered,  has no strong finite size effects already at $N=256$ spins (c$'''$ and c$''''$).

As we already mentioned, in Eqs. (\ref{eq:H}-\ref{eq:J}) the number of targets $P$ plays the same role of the number of memories to be stored into a Hebbian matrix in the AGS theory  \cite{amit1985storing,amit1987statistical} predicting  the impossibility of memory retrieval, i. e., the inability to retrieve the encoded $\xi$ states,  for  $\alpha$ larger than a given critical value  $\alpha_c$.
This qualitative change in the degree of complexity of our system can be demonstrated in Fig. \ref{OVERLAP}, where we report the temperature behavior of the probability density function $P(q)$  of the overlap
\begin{equation}
q_{ab}\equiv \frac{1}{N}\sum_{k=1}^NS_k^{(a)}S_k^{(b)}
\label{eq:q}
\end{equation}
between couples of replicas $\bm S^{(a)}$ and $\bm S^{(b)}$. At high temperature each equilibrium configuration, will be uncorrelated from the others, because of strong thermal fluctuations. For low  $T$, at $\alpha\simeq 0$, that is $P/N\to 0$ as $N\to\infty$, the model dynamics will tend to align, or counter-align, to the $\xi$ patterns, as mentioned in Eq. \ref{I_simplified} for the generalized Mattis model. The values of the overlap between such states will, then, converge to only two possible values as $N$ increases, one the inverse of the other, and the distribution will display two distinct symmetric peaks.  As the number of patterns to be satisfied increases with the number of micromirrors, though, frustration arises as it becomes more and more  unfeasible to satisfy any of them with a configurations of $\bm S$ minimizing the Hamiltonian $\mathcal H[\bm S]$, cf. Eq. (\ref{eq:H}). The change of the $P(q)$ from a low temperature two peak distribution into a multi peaks distribution is a clear evidence of the onset of such frustration and the relative complexity in the equilibrium state organization.  We stress that in Fig. \ref{OVERLAP} the overlap distributions displayed are for a single realization of the random couplings.
Moreover, from these curves we extracted the values of the $T_G$ (see methods), and these are reported in Fig. \ref{OVERLAP}e as a function of $\alpha$ for  $N=225$ (green dots fitted with green curve). The curves represent the fit with the model from phase transition equation from replica symmetry breaking calculation ( eq. \ref{Phasetranstion}) leaving $A$ as a free parameters. By fitting the data for $T_G$ with the model we retrieve a constant $A$ of 1.44$\pm$ 0.6, thus resulting within the fit error with the theoretically predicted value ($A$=1).

\begin{figure}[h!]
\centering
\includegraphics[width=14 cm]{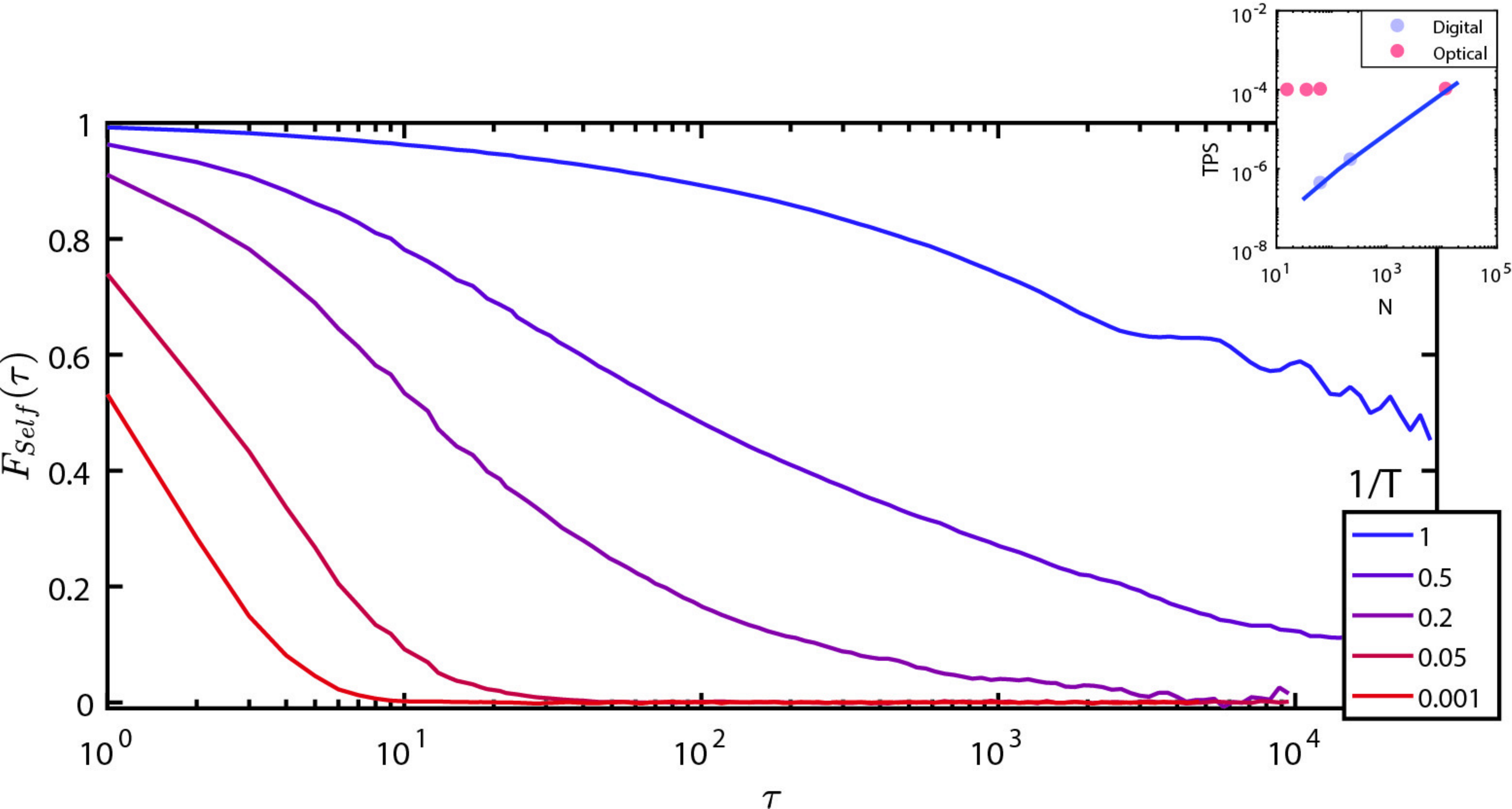}
\caption{Self-correlation function for a system with $N=12100= 110 \times 110$ spins and $P= 1369$ targets. This has been obtained with a total of $0.7$ billions flipping individual mirror flipping operations, performed in $20$ hours. The inset shows the timing behavior (Time Per Step TPS) as function of $N$ for the optical (red dots) and digital (blue dots) simulations. \label{Long_Measure}}
\end{figure}

In order to test our optical procedure we performed numerical experiments with a few spins, from $N=16$ to $N=225$, however the real advantage of the proposed method comes about when the sample size and the number of targets is large.
In Fig. \ref{Long_Measure}  we report a measurement with a  large fully connected spin glass ($N=12100$  micromirror-combined spins organized into a $110\times 110$ 2D square lattice) with $\alpha =0.113$.  We retrieve a time per step (TPS) of $\simeq 10^{-4}$ s/step for this large optical experiment. This TPS turns out to be only barely dependent on $N$, as we show in the
inset of Fig. \ref{Long_Measure} in which we report the TPS for the Optical simulation as red circles. The time per step for the digital simulation is, instead,  growing with $N$ as expected (inset of Fig. \ref{Long_Measure} blue circles). The optical simulation is thus outperforming the digital one  at $N \sim 12000$. We note that this configuration is exploiting only a small portion of the full DMD size ($48400$ mirrors out of $786432$, i.e., one over $16$), thus potentially still providing a further advantage with respect to the fully digital approach for simulations with larger $N$.


In conclusion, we demonstrate the possibility to simulate the dynamics of a system with $N$ spins and $N^2$ couplings with random couplings and a low temperature spin-glass phase. Our approach has the advantage of
(i)  being scalable: the possibility to simulate very large systems without affecting the calculation speed, because the instantaneous interference enables to access the energy difference without requiring to individually calculate all the coupling terms.
(ii) varying the constraints density so as to survey both a ferromagnetic-like ordering, retrieving the transmission matrix patterns, and a spin-glass freezing, i.e., loss of memory,
 by playing with the number of optical targets in which the intensity is monitored,
(iii) vary the dynamic variables of the system, for Ising spins, needing each one four micro-mirrors to be constructed, both to multiple discrete states variables and to complex continuous phasors.
(iiii)The same approach may be employed with multi state spins, employing higher dimensional superpixel capable to generate many different phase and amplitude values. In this configuration, as far as the number of targets per spin is below the critical $\alpha_c(T)$  value of memory retrieval, the analysis of the ferromagnetic Mattis-like states allows to reconstruct the transmission matrix elements of the random opaque medium through the states $\bm S$ visited in the optical simulation, as {\em memories} learned in a neural network.


\appendix

\section{Appendixes}

\subsection*{Random couplings and transmission matrix}
In Eq. (\ref{eq:J}) the couplings have been defined as
\begin{equation}
J_{ij}=\frac{1}{N}\sum_{\nu=1}^P v_{ij}^{(\nu)} \quad ; \quad  v_{ij}^{(\nu)}\equiv \xi^{(\nu)}_{i}\bar \xi^{(\nu)}_{j} \quad ;\quad  \xi^{(\nu)}_{i}=\rho^{(\nu)}_{i}+\imath\upsilon^{(\nu)}_{i} \in \mathbb{C}
\label{J_}
\end{equation}
where the transmission matrix properties \cite{beenakker1997random} show that real and imaginary parts of $\xi$ are both Gaussian distributed as ($x=\rho,\upsilon$),
\begin{equation}
P(x)=\frac{1}{\sqrt{2\pi\sigma^2}}e^{-\frac{x^2}{2\sigma^2}} \ .
\label{P_xi}
\end{equation}
The $J_{ij}$ matrix is Hermitian because $ \xi^{(\nu)}_{i}\bar \xi^{(\nu)}_{j} = {\overline{ \xi^{(\nu)}_{j}\bar \xi^{(\nu)}_{i}}}$.
The intensity per mirror $I/N$  is then, expressed as:
\begin{equation}
\frac{I}{N}=\sum^N_{ij} J_{ij}S_iS_j=\sum^N_{j>i} (J_{ij}+J_{ji}) S_i S_j +\sum^N_{i} J_{ii} =  \sum^N_{i \neq j} J^R_{ij} S_i S_j + \mbox{const} \ .
\label{I_full}
\end{equation}
Note that even if $J_{ij}$, $i\neq j$, is in general complex valued, each term of the sum $J_{ij}+J_{ji} = 2 J^R_{ij} $ is real and symmetric in $i\leftrightarrow j$ because it results form $\xi_{i}\bar\xi_{j} +\xi_{j}\bar\xi_{i}$.
For a  large number of targets $P$, as in the case $P=\alpha N$ with an $\alpha=O(1)$, the $J_{ij}$ element is the sum of many random numbers (each one the product of two Gaussian random numbers) and, for the central limit theorem, will be Gaussian distributed.

For a single target $P=1$ to derive the probability distribution of the values of $J_{ij}=\xi_i\bar\xi_j/N$  we can, instead,  integrate the expression (we discard the index $\nu$)
\begin{eqnarray}
\label{P_J_diag}
P(J_{ii})=\int \frac{d\rho ~d\upsilon}{2\pi\sigma^2}
e^{-\frac{\rho^2+\upsilon^2}{2\sigma^2}}
\delta \left( J_{ii}N-\rho^2-\upsilon^2 \right)
= \frac{N}{2\sigma^2} e^{-\frac{J_{ii}N}{2\sigma^2}} \ ,
\end{eqnarray}
for the diagonal entries $J_{ii}=|\xi_i|^2$, while for $i\neq j$  the distribution of $J^R_{ij}=\rho_i\rho_j+\upsilon_i\upsilon_j$ reads
\begin{eqnarray}
\nonumber
P(J^R_{ij})&=&\int \frac{d\rho_i ~d\upsilon_i}{2\pi\sigma^2}\frac{d\rho_j ~d\upsilon_j}{2\pi\sigma^2}
e^{-\frac{\rho_i^2+\upsilon_i^2+\rho_j^2+\upsilon_j^2}{2\sigma^2}}
\delta \left( J^R_{ij}N-\rho_i\rho_j-\upsilon_i\upsilon_j\right)
\\
&=&
\frac{N}{\sqrt{4\pi\sigma^4}}\int \frac{dz}{\sqrt{z}}\exp\left\{-z-\frac{N^2\left(J^R_{ij}\right)^2}{4 \sigma^4}\frac{1}{ z}\right\}
= \frac{N}{2\sigma^2} e^{- \frac{\left|J^R_{ij}\right| N}{\sigma^2}} \ ,
\label{P_J_off}
\end{eqnarray}
where we have used the relationship
\begin{eqnarray}
\nonumber
{\mathcal I}(\gamma) \equiv \int_0^\infty \frac{dz}{\sqrt{z}} e^{-z-\frac{\gamma^2}{z}} = \frac{\sqrt{\pi}}{e^{2|\gamma|}}  \ .
\end{eqnarray}

To verify that the distributions are, indeed, exponential, thus indirectly confirming the Gaussian distribution of the transmission matrix elements  $\xi$, and to estimate their variance by interpolation, we can experimentally measure the distribution of the $J's$, or, equivalently, of the $v_{ij}=J_{ij}N$, cf. Eq. (\ref{eq:J}) with $P=1$.
We resorted to a pair of measurements of the propagation of the signal from two superpixels $S_1$, $S_2$ on a single target.
In a first run we turn on a single superpixel and measure the diagonal contributions independently:  the intensity at the target $k=1,2$ is the squared modulus  of the  amplitude transmitted by the superpixel, thus,
\begin{equation}
I_{{\rm diag},k} \equiv v_{kk}=\xi_{k}\bar \xi_{k}=\rho_k^2+\upsilon_k^2,
\label{I_D}
\end{equation}
In the second run we turn on the couple of superpixels  $S_1=S_2=1$ and measure the overall intensity at the target $1$
\begin{eqnarray}
\nonumber
I_{12}[S_1=1,S_2=1]=v_{11}+v_{12}+v_{21}+v_{22} &=&
\xi_{1}\bar \xi_{1}+\xi_{1}\bar \xi_{2}+\xi_{2}\bar \xi_{1}+\xi_{2}\bar \xi_{2}
\\
\nonumber
&=&\rho_1^2+\upsilon_1^2 +\rho_2^2+\upsilon_2^2 + 2\rho_1\rho_2+2\upsilon_1\upsilon_2
\end{eqnarray}
 whose extra,  off-diagonal contribution, is
\begin{eqnarray}
I_{\rm off~diag} \equiv I_{12} - I_{{\rm diag},1}- I_{{\rm diag},2}
= \xi_1 \bar \xi_2 +  \bar \xi_1  \xi_2 = 2\rho_1\rho_2+2\upsilon_1\upsilon_2 =  v_{12}+v_{21} = 2 v^R_{12} = 2N J_{12}^R   \ .
\end{eqnarray}

Measuring several values of  $I_{12} $ and $I_{{\rm diag},k}$  we find that both $I_{\rm off~diag}$ and $I_{\rm diag,k}$ have an exponential distribution as shown in the inset of Fig. \ref{fig:setup}. Looking in semi-log scale and fitting with the interpolating function $A~ e^{-B J}$  the slope of the distribution $P(J_{ij}^R)$ turns out to be twice the slope of $P(J_{ii})$, as predicted by Eqs. (\ref{P_J_diag},\ref{P_J_off}).

\subsection*{Superpixel method}
Light reflected by the DMD produces a diffraction pattern composed by a square array of intensity maxima. These maxima correspond to maxima of interference due to in-phase summation of all the contribution from the each light ray reflected by pixels.  By placing a lens (L1 with focal lenght $f$ = 200 mm) to collect DMD light we have the zeroth order fringe appears at coordinates $[x_f,y_f]= [0,0]$ and the first order at coordinates $[x_f,y_f]= [a,a]$ with $a$ the lattice primitive vector $a=\frac{\lambda f}{d}$, where $\lambda$ is the wavelength of light (0.532 nm), and $d$ is the DMD pixel Pitch ( 13.68 $\mu m$).

As shown in [Opt Expr 22 1364] it is possible to exploit a DMD device to control both amplitude and phase of coherent light beam. This technique is based on the phase delays accumulated by light rays corresponding to different pixels, if light is collected only from special positions in Fourier space. The first step is to collect light thought a spatial filter (iris $I$) which is placed off at coordinates $[x_f,y_f]= [a/2,a/2]$ in the Fourier space. In such case each DMD pixel acquires a relative phase delay wick depends on its position in the real space: in particular, if pixel are identified by a couple of integers  $[x_i, y_i]$, pixels with even $T=x_i + y_i$ contribute with a +1 pre-fator, while , pixels with odd $T$ contribute with a -1 pre-factor.
The we organize the DMD into superpixels composed by 2x2 DMD pixels. In order to merge each pixel into a single contribution we reduce the optical resolution in order to ``blur them together'' this is performed thanks to the iris  $I$ which is tuned on a diameter of 2.66 mm. This diameter produces an optical resolution of 32 $\mu m$ thus blurring together four pixels.

In order to ensure the validity of the proposed model we also realized a system in which the size of the superpixel matches the size of the disorder grain in the scattering medium. This is obtained employing a ground glass diffuser (``DG10-220'' thorlabs with grain size of 63 $\mu m$) and an optical system with magnification 2, producing a superpixel image of 64 $\mu m$ side.

\subsection*{Overlap $q$, $P(q)$ and calculation of  $T_G$ }
In the fast protocol (Fig. \ref{OVERLAP}a-b-c) $P(q)$ for a given  [ $\beta$ , $\alpha$] configuration, has been obtained by performing $50$ fast (25 sweeps)
simulations of replicas of the system  (each starting from a random initial configuration) and calculating a value of $q$ as the scalar product between the final states for each pair of replicas. $P(q)$ is obtained by extracting the occurrence probability for each $q$ value. Even if this approach provides states not completely thermalized, the comparison with longer numerical simulations  (Fig. \ref{OVERLAP}d) demonstrates a good agreement about the onset of the spin-glass phase.
In the long protocol (Fig. \ref{OVERLAP}d) $P(q)$ in a given $\beta$ and  $\alpha=1$ configuration, has been obtained by monitoring each replica for a long time (8000 Monte Carlo sweeps in the Optical Spin Glass simulation). Then $P(q)$ has been obtained as the normalized histogram of the scalar product of  two replicas at the same time.

A first raw estimate of $T_G$ has been extracted by calculating at the  Kurtosis ${K}(\beta)$ of the $P(q)$  distribution for various values of $\alpha$ as the temperature is decreased:
\begin{equation}
{K}(T) = \frac{\langle\left( q-\langle q\rangle\right)^4\rangle}{\langle \left( q-\langle q \rangle\right)^2\rangle^2}
\end{equation}
where the average is taken over the distribution $P(q)$. We underline that the $P(q)=P_\xi(q)$ considered here are for a given realization of the disorder (transmission patterns $\xi$, or, equivalently, spin couplings $J$).
We estimate $T_G(N)$, at a given system size, as the temperature at which ${K}(T)$ is sensitively less than $3$, i. e., $P(q)$ is no more a high temprerature Gaussian.

\subsection*{Numerical simulations}

To check the performances of the OSG we also have performed standard numerical simulations. The implemented dynamics is a standard Metropolis Monte Carlo, as in the optical spin-glass simulation. We built the random generator  for the coupling factors $\xi$, rather than for $J$, in order to ensure correspondence with the optical simulations, generating $M$ $N$-components complex vectors $\xi^{(1)},\xi^{(2)},\dots,\xi^{(M)}$, which represent the transmission from the superpixel on to the targets in  the optical setup.  Subsequently, the $J_{ij}$ element of the coupling matrix is calculated from those vectors as follows:
\begin{equation}
J_{ij} = \frac{1}{2}\sum_{\nu=1}^M \left( \xi^{(\nu)}_i \bar\xi^{(\nu)}_j + \mbox{  c.c.} \right)=
\sum_{\nu=1}^M \left(\rho_i^{(\nu)}\rho_j^{(\nu)}+\upsilon^{(\nu)}_i\upsilon_j^{(\nu)}\right)
\end{equation}
During the dynamics, for the energy difference at each step, we only perform the sum over the couples which include the spin that have been flipped at that step; in such a way that we only need to sum $O(N)$ terms instead of all possible $O(N^2)$ terms. The final formula for the energy difference is therefore, assuming the flipped spin to be the $\bar \i$-th one:
\begin{equation}
\Delta H = \sigma_{\,\bar \i} \sum_{j} J_{\,\bar \i, j} \,\sigma_j.
\end{equation}

The simulation have been performed by a multi core  cluster composed by
\begin{itemize}
\item \#3 nodes each composed by with \#2 CPU XEON E7-2860 (20 threads per CPU) clocked at 2.27GHz with 512 GB ram per CPU.
\item \#6 nodes each composed by with \#2 CPU XEON E5-2630 (12 threads per CPU) clocked at 2.30GHz with 72 GB ram per CPU.
\end{itemize}

\subsection*{Replica theory of the complex Gaussian Hopfield model}

Defining the  extensive ``energy" of a superpixel configuration ${\bm s}$ as $\mathcal H[\bm s] = -I/2/N = O(N)$,
the statistical mechanical model for our optical system displays the Hopfield-like Hamiltonian
\bea
\mH[\bs] &=& -\frac{1}{2}\sum_{i\neq j} J_{ij} s_is_j -\sum_{i=1}^N s_i H_i
\eea
with
\bea
J_{ij} &\equiv&  \frac{1}{N}\sum_{\mu =1}^p \Re[\ximu_i \bximu_j]
\\
H_i&\equiv& \frac{1}{\sqrt{N}}\sum_{\mu=1}^p
\Re[ \bar h^{(\mu)} \xi_i^{(\mu)} ]
= \sum_{\mu=1}^p
 \Re\left[\left(
\frac{ h_0^{(\mu)}}{\sqrt{N}} + \frac{1}{N}\sum_{j=1}^N\bximu_j
 \right) \ximu_i
\right]
\eea
where the local fields $1/N \sum_{j=1}^N\bximu_j$ arise in the superpixel composition but their overall contribution to the Hamiltonian is, actually, of $O(1)$ and, thus negligible with respect to the extensive contribution of the pairwise spin-interaction. The fields $h_0^{(\mu)}$ can be introduced to identify  Mattis states, i.e., those states most aligned along a given transmission pattern $\xi^{(\mu)}$, and are to be sent to zero eventually, in order to be able to study also the ferromagnetic transition
at low enough  $\alpha$. We focus in this work to present the spin-glass behavior and we will present the analysis of Mattis states elsewhere. Therefore, we set  $h_0^{(\mu)}=0$, $\forall \mu$ in the following.

To compute the average of the free energy over the distribution of the transmission matrix complex-valued elements we need to replicate the system, i. e.,
\bea
\beta \Phi &=& - \lim_{N\to \infty} \frac{1}{N}\mathbb{E}\left[\ln Z\right]_\xi
\\
\nn
&=& \lim_{n\to 0} \lim_{N\to \infty} \frac{\mathbb{E}\left[ Z^n \right]_\xi-1}{n N}
\eea
where the replicated partition function for a given random transmission (given transmission matrix) reads
\bea
\nn
Z^n&=&(N\beta)^{p/2} \int \prod_{\mu a} \frac{dr_a^{\mu}}{\sqrt{2\pi}} \frac{dt_a^{\mu}}{\sqrt{2\pi}} e^{-\frac{N\beta}{2}\sum_{\mu a}\left[\left(r_a^{(\mu)}\right)^2+\left(t_a^{(\mu)}\right)^2\right]}
 \sum_{\{s\}} e^{F\left(\{ \underline \xi \}, \{ \vec s \},{\vec {\underline m }}\right)}
\eea
with
\bea
F&\equiv& \beta \sum_{a \mu}
\left[
r_a^{(\mu)} \sum_i\ximu_{R,i} \ssa_i +t_a^{(\mu)} \sum_i\ximu_{I,i} \ssa_i \right]+ O(1)  \ .
\eea
After having averaged over the transmission matrix elements distribution, one obtains
\bea
\nn
\mathbb{E}\left[ Z^n \right]_\xi &=& (N\beta)^{p/2} \int \prod_{\mu a} \frac{dr_a^{\mu}}{\sqrt{2\pi}} \frac{dt_a^{\mu}}{\sqrt{2\pi}} e^{-\frac{N\beta}{2}\sum_{\mu a}\left[\left(r_a^{(\mu)}\right)^2+\left(t_a^{(\mu)}\right)^2\right]}
 \sum_{\{\vec s\}} \mathbb{E}\left[ e^{F\left(\{ \underline \xi \}, \{ \vec s \},{\vec {\underline m }}\right)} \right]_\xi
\eea
with
\bea
\nn
\mathbb{E}\left[ e^{F} \right]_\xi &=&\prod_{\mu i} e^{
\frac{\beta^2\sigma^2}{2}\sum_{a b}\left(r^{(\mu)}_ar^{(\mu)}_b+t^{(\mu)}_at^{(\mu)}_b\right)\ssa_i\ssb_i}  \ .
\eea
Let us now define the vector of the spin state through the $n$ replicas with an arrow $\vec s\equiv \{s_1, \ldots, s_n\}$, to distinguish it from the configuration of all spins in a replica that we represent by a bold $\bs$.
Properly rescaling some integration variables, defining the replica overlap
\bea
q_{ab}&\equiv& \frac{1}{N}\sum_i \ssa_i \ssb_i \ ,
\eea
introducing  the matrix
\bea
\mM_{ab} &\equiv& \delta_{ab}(1-\gamma)-\gamma q_{ab}
\eea
and defining the rescaled inverse temperature $\gamma\equiv  \beta \s^2$
we obtain the average replicated partition function

\bea
\mathbb{E}\left[ Z^n \right]_\xi & =& (\beta N)^{p(1-n)} \int \prod_{a<b} dq_{ab}  \frac{d\la_{ab}}{2\pi \imath} e^{-nN\mS[q,\la]}
\label{FEN}
\\
\nn
\mS[q,\la]&\equiv& \frac{1}{n}\sum_{a<b}q_{ab}\la_{ab}-\frac{1}{n}\ln \mZ[\la]+\frac{\alpha}{n}\ln \det \mM[q]
\\
\mZ[\la]&\equiv&  \sum_{\vec s}e^{H[\la,\vec s]}
\nn
\\
H[\la,\vec s]&\equiv& \sum_{a<b}\la_{ab}s_a s_b
\nn
\eea
The free energy is obtained by computing the
saddle point of the integral (\ref{FEN}) for $N\gg 1$, yielding
\bea\mathbb{E}\left[ Z^n \right]_\xi & \simeq&    e^{nN\mS[q_{\rm sp},\la_{\rm sp}]}+O\left(\frac{1}{N}\right)
\nn
\\
\beta \Phi &=& - \lim_{N\to \infty} \frac{1}{N}\mathbb{E}\left[\ln Z\right]_\xi \simeq  \lim_{n\to 0} \lim_{N\to \infty} \frac{\mathbb{E}\left[ Z^n \right]_\xi-1}{n N}
=\mS[q_{\rm sp},\la_{\rm sp}]
\label{eq:S}
\eea
The saddle point equations, i. e., the self-consistency equations for the matrix {\it order parameters} $q_{ab}$ and $ \la_{ab}$, read
\bea
\la_{ab}&=& 2\alpha\gamma \left(\mM^{-1}
\right)_{ab}
\label{eq:la_ab}
\\
q_{ab}&=&\frac{ \sum_{\{\vec s\}}s_as_b~ e^{H[\la,\vec s]}}{ \sum_{\{\vec s\}}e^{H[\la,\vec s]}} = \lan s_as_b\ran
\label{eq:q_ab2}
\eea

\subsection{Replica Symmetry}
A first solution, certainly valid in the paramagnetic phase, can be put forward assuming replica symmetry, that is, assuming the overlap matrix form
\bea
q_{ab } &=& (1-\delta_{ab}) q
\eea
and, consequently,
\bea
\mM_{ab} &=& (1-\gamma+\gamma q) \delta_{ab}-\gamma q , \qquad {a\neq b}
\eea
The saddle point equations of the RS solution read
\bea
\lambda&=&\frac{2\alpha \gamma^2 q}{(1-\gamma \chi)^2}
\label{eq:la}
\\
q&=&\int \frac{dz}{\sqrt{2\pi}}e^{-z^2/2} \tanh^2\left(\sqrt{\la} z\right)
\label{eq:q2}
\eea
with $\chi\equiv 1-q$, the replica symmetric free energy is
\bea
\beta \Phi &=& \frac{\lambda}{2}(1-q)-\int  \frac{dz}{\sqrt{2\pi}}e^{-z^2/2} \ln2\cosh\left(\sqrt{\la} z\right)
+\alpha\ln(1-\gamma\chi)-\alpha\gamma \frac{q}{1-\gamma\chi}
\eea
In particular, the  paramagnetic free energy will be ($q=0$)
\bea
\beta\Phi_{\rm PM} = -\ln 2 +\alpha \ln(1-\gamma)
\label{eq:F_PM}
\eea
\subsection{Stability and phase transition}
The stability eigenvalues of the replicated free energy in the RS Ansatz are three: the so-called replicon eigenvalue
\bea
\Lambda_1&=&
=  \frac{1}{2\alpha} \left(
\frac{1-\gamma\chi}{\gamma}
\right)^2
- \int \frac{dz}{\sqrt{2\pi}}e^{-z^2/2} \left[1-\tanh^2\left(\sqrt{\la} z\right)\right]^2
\eea
and the longitudinal and the anomalous eigenvalues, that coincide for $n\to 0$:
\bea
\Lambda_{2,3}=  \frac{1}{2\alpha\gamma^2} \left[\left(
1-\gamma
\right)^2+4q \gamma(1-\gamma)+3q^2\gamma^2\right]
-(1-4q+3r)
\eea

On the paramagnetic phase ($q=0$) the stability conditions become one
$\Lambda_1=\Lambda_{2,3}\geq 0$:
\bea
\frac{T}{\sigma^2} > 1 + \sqrt{2\alpha}.
\label{Phasetranstion}
\eea
that is the critical temperature of the transition from the paramagnetic to the spin-glass phase, that depends on  the ratio  $\alpha$
between targets and spins.
The phase transition turns out to be continuous in the order paramater $q$.

\subsection{Aknowledgements}
M.L. thanks  ``Fondazione CON IL SUD'',  ``Grant Brains2south, Project ``Localitis''.
Manuscript submitted to PNAS.



\bibliography{OCSGDTC}

\begin{thebibliography}{38}
\expandafter\ifx\csname natexlab\endcsname\relax\def\natexlab#1{#1}\fi
\expandafter\ifx\csname bibnamefont\endcsname\relax
  \def\bibnamefont#1{#1}\fi
\expandafter\ifx\csname bibfnamefont\endcsname\relax
  \def\bibfnamefont#1{#1}\fi
\expandafter\ifx\csname citenamefont\endcsname\relax
  \def\citenamefont#1{#1}\fi
\expandafter\ifx\csname url\endcsname\relax
  \def\url#1{\texttt{#1}}\fi
\expandafter\ifx\csname urlprefix\endcsname\relax\def\urlprefix{URL }\fi
\providecommand{\bibinfo}[2]{#2}
\providecommand{\eprint}[2][]{\url{#2}}

\bibitem[{\citenamefont{Harris et~al.}(2017)\citenamefont{Harris, Steinbrecher,
  Prabhu, Lahini, Mower, Bunandar, Chen, Wong, Baehr-Jones, Hochberg
  et~al.}}]{harris2017quantum}
\bibinfo{author}{\bibfnamefont{N.~C.} \bibnamefont{Harris}},
  \bibinfo{author}{\bibfnamefont{G.~R.} \bibnamefont{Steinbrecher}},
  \bibinfo{author}{\bibfnamefont{M.}~\bibnamefont{Prabhu}},
  \bibinfo{author}{\bibfnamefont{Y.}~\bibnamefont{Lahini}},
  \bibinfo{author}{\bibfnamefont{J.}~\bibnamefont{Mower}},
  \bibinfo{author}{\bibfnamefont{D.}~\bibnamefont{Bunandar}},
  \bibinfo{author}{\bibfnamefont{C.}~\bibnamefont{Chen}},
  \bibinfo{author}{\bibfnamefont{F.~N.} \bibnamefont{Wong}},
  \bibinfo{author}{\bibfnamefont{T.}~\bibnamefont{Baehr-Jones}},
  \bibinfo{author}{\bibfnamefont{M.}~\bibnamefont{Hochberg}},
  \bibnamefont{et~al.}, \bibinfo{journal}{Nature Photonics}
  \textbf{\bibinfo{volume}{11}}, \bibinfo{pages}{447} (\bibinfo{year}{2017}).

\bibitem[{\citenamefont{Sparrow et~al.}(2018)\citenamefont{Sparrow,
  Martin-Lopez, Maraviglia, Neville, Harrold, Carolan, Joglekar, Hashimoto,
  Matsuda, O~Brien et~al.}}]{sparrow2018simulating}
\bibinfo{author}{\bibfnamefont{C.}~\bibnamefont{Sparrow}},
  \bibinfo{author}{\bibfnamefont{E.}~\bibnamefont{Martin-Lopez}},
  \bibinfo{author}{\bibfnamefont{N.}~\bibnamefont{Maraviglia}},
  \bibinfo{author}{\bibfnamefont{A.}~\bibnamefont{Neville}},
  \bibinfo{author}{\bibfnamefont{C.}~\bibnamefont{Harrold}},
  \bibinfo{author}{\bibfnamefont{J.}~\bibnamefont{Carolan}},
  \bibinfo{author}{\bibfnamefont{Y.~N.} \bibnamefont{Joglekar}},
  \bibinfo{author}{\bibfnamefont{T.}~\bibnamefont{Hashimoto}},
  \bibinfo{author}{\bibfnamefont{N.}~\bibnamefont{Matsuda}},
  \bibinfo{author}{\bibfnamefont{J.~L.} \bibnamefont{O~Brien}},
  \bibnamefont{et~al.}, \bibinfo{journal}{Nature}
  \textbf{\bibinfo{volume}{557}}, \bibinfo{pages}{660} (\bibinfo{year}{2018}).

\bibitem[{\citenamefont{Lin et~al.}(2018)\citenamefont{Lin, Rivenson, Yardimci,
  Veli, Luo, Jarrahi, and Ozcan}}]{lin2018all}
\bibinfo{author}{\bibfnamefont{X.}~\bibnamefont{Lin}},
  \bibinfo{author}{\bibfnamefont{Y.}~\bibnamefont{Rivenson}},
  \bibinfo{author}{\bibfnamefont{N.~T.} \bibnamefont{Yardimci}},
  \bibinfo{author}{\bibfnamefont{M.}~\bibnamefont{Veli}},
  \bibinfo{author}{\bibfnamefont{Y.}~\bibnamefont{Luo}},
  \bibinfo{author}{\bibfnamefont{M.}~\bibnamefont{Jarrahi}}, \bibnamefont{and}
  \bibinfo{author}{\bibfnamefont{A.}~\bibnamefont{Ozcan}},
  \bibinfo{journal}{Science} \textbf{\bibinfo{volume}{361}},
  \bibinfo{pages}{1004} (\bibinfo{year}{2018}).

\bibitem[{\citenamefont{Peruzzo et~al.}(2014)\citenamefont{Peruzzo, McClean,
  Alan, and O~Brien}}]{peruzzo2014variational}
\bibinfo{author}{\bibfnamefont{A.}~\bibnamefont{Peruzzo}},
  \bibinfo{author}{\bibfnamefont{J.}~\bibnamefont{McClean}},
  \bibinfo{author}{\bibfnamefont{S.}~\bibnamefont{Alan}}, \bibnamefont{and}
  \bibinfo{author}{\bibfnamefont{J.~L.} \bibnamefont{O~Brien}},
  \bibinfo{journal}{Nature communications} \textbf{\bibinfo{volume}{5}},
  \bibinfo{pages}{4213} (\bibinfo{year}{2014}).

\bibitem[{\citenamefont{Roques-Carmes et~al.}(2019)\citenamefont{Roques-Carmes,
  Shen, Zanoci, Prabhu, Atieh, Jing, Dub\v{c}ek, \v{C}eperi\'{c}, Joannopoulos,
  Englund et~al.}}]{Roques-Carmes:19}
\bibinfo{author}{\bibfnamefont{C.}~\bibnamefont{Roques-Carmes}},
  \bibinfo{author}{\bibfnamefont{Y.}~\bibnamefont{Shen}},
  \bibinfo{author}{\bibfnamefont{C.}~\bibnamefont{Zanoci}},
  \bibinfo{author}{\bibfnamefont{M.}~\bibnamefont{Prabhu}},
  \bibinfo{author}{\bibfnamefont{F.}~\bibnamefont{Atieh}},
  \bibinfo{author}{\bibfnamefont{L.}~\bibnamefont{Jing}},
  \bibinfo{author}{\bibfnamefont{T.}~\bibnamefont{Dub\v{c}ek}},
  \bibinfo{author}{\bibfnamefont{V.}~\bibnamefont{\v{C}eperi\'{c}}},
  \bibinfo{author}{\bibfnamefont{J.~D.} \bibnamefont{Joannopoulos}},
  \bibinfo{author}{\bibfnamefont{D.}~\bibnamefont{Englund}},
  \bibnamefont{et~al.}, \bibinfo{journal}{Conference on Lasers and
  Electro-Optics} p. \bibinfo{pages}{FTu4C.2} (\bibinfo{year}{2019}).

\bibitem[{\citenamefont{Farhat et~al.}(1985)\citenamefont{Farhat, Psaltis,
  Prata, and Paek}}]{farhat1985optical}
\bibinfo{author}{\bibfnamefont{N.~H.} \bibnamefont{Farhat}},
  \bibinfo{author}{\bibfnamefont{D.}~\bibnamefont{Psaltis}},
  \bibinfo{author}{\bibfnamefont{A.}~\bibnamefont{Prata}}, \bibnamefont{and}
  \bibinfo{author}{\bibfnamefont{E.}~\bibnamefont{Paek}},
  \bibinfo{journal}{Applied optics} \textbf{\bibinfo{volume}{24}},
  \bibinfo{pages}{1469} (\bibinfo{year}{1985}).

\bibitem[{\citenamefont{Edwards and Anderson}(1975)}]{edwards1975}
\bibinfo{author}{\bibfnamefont{S.}~\bibnamefont{Edwards}} \bibnamefont{and}
  \bibinfo{author}{\bibfnamefont{P.}~\bibnamefont{Anderson}},
  \bibinfo{journal}{J. Phys. Lett.} \textbf{\bibinfo{volume}{5}},
  \bibinfo{pages}{965} (\bibinfo{year}{1975}).

\bibitem[{\citenamefont{M{\'e}zard et~al.}(1987)\citenamefont{M{\'e}zard,
  Parisi, and Virasoro}}]{mezard1987spin}
\bibinfo{author}{\bibfnamefont{M.}~\bibnamefont{M{\'e}zard}},
  \bibinfo{author}{\bibfnamefont{G.}~\bibnamefont{Parisi}}, \bibnamefont{and}
  \bibinfo{author}{\bibfnamefont{M.}~\bibnamefont{Virasoro}},
  \emph{\bibinfo{title}{Spin glass theory and beyond: An Introduction to the
  Replica Method and Its Applications}}, vol.~\bibinfo{volume}{9}
  (\bibinfo{publisher}{World Scientific Publishing Company},
  \bibinfo{year}{1987}).

\bibitem[{\citenamefont{Young}(1998)}]{young1998spin}
\bibinfo{author}{\bibfnamefont{A.~P.} \bibnamefont{Young}},
  \emph{\bibinfo{title}{Spin glasses and random fields}},
  vol.~\bibinfo{volume}{12} (\bibinfo{publisher}{World Scientific},
  \bibinfo{year}{1998}).

\bibitem[{\citenamefont{Fisher and Huse}(1988)}]{fisher1988equilibrium}
\bibinfo{author}{\bibfnamefont{D.~S.} \bibnamefont{Fisher}} \bibnamefont{and}
  \bibinfo{author}{\bibfnamefont{D.~A.} \bibnamefont{Huse}},
  \bibinfo{journal}{Phys. Rev. B} \textbf{\bibinfo{volume}{38}},
  \bibinfo{pages}{386} (\bibinfo{year}{1988}).

\bibitem[{\citenamefont{Leuzzi et~al.}(2009)\citenamefont{Leuzzi, Parisi,
  Ricci-Tersenghi, and Ruiz-Lorenzo}}]{leuzzi2009ising}
\bibinfo{author}{\bibfnamefont{L.}~\bibnamefont{Leuzzi}},
  \bibinfo{author}{\bibfnamefont{G.}~\bibnamefont{Parisi}},
  \bibinfo{author}{\bibfnamefont{F.}~\bibnamefont{Ricci-Tersenghi}},
  \bibnamefont{and} \bibinfo{author}{\bibfnamefont{J.~J.}
  \bibnamefont{Ruiz-Lorenzo}}, \bibinfo{journal}{Phys. Rev. Lett.}
  \textbf{\bibinfo{volume}{103}}, \bibinfo{pages}{267201}
  (\bibinfo{year}{2009}).

\bibitem[{\citenamefont{Temesvari}(2010)}]{temesvari2010theising}
\bibinfo{author}{\bibfnamefont{T.}~\bibnamefont{Temesvari}},
  \bibinfo{journal}{Nucl. Phys. B} \textbf{\bibinfo{volume}{829}},
  \bibinfo{pages}{534} (\bibinfo{year}{2010}).

\bibitem[{\citenamefont{Baity-Jesi et~al.}(2018)\citenamefont{Baity-Jesi,
  Calore, Cruz, Fernandez, Gil-Narvion, Gordillo-Guerrero, I\~niguez, Maiorano,
  Marinari, Martin-Mayor et~al.}}]{janus2018aging}
\bibinfo{author}{\bibfnamefont{M.}~\bibnamefont{Baity-Jesi}},
  \bibinfo{author}{\bibfnamefont{E.}~\bibnamefont{Calore}},
  \bibinfo{author}{\bibfnamefont{A.}~\bibnamefont{Cruz}},
  \bibinfo{author}{\bibfnamefont{L.~A.} \bibnamefont{Fernandez}},
  \bibinfo{author}{\bibfnamefont{J.~M.} \bibnamefont{Gil-Narvion}},
  \bibinfo{author}{\bibfnamefont{A.}~\bibnamefont{Gordillo-Guerrero}},
  \bibinfo{author}{\bibfnamefont{D.}~\bibnamefont{I\~niguez}},
  \bibinfo{author}{\bibfnamefont{A.}~\bibnamefont{Maiorano}},
  \bibinfo{author}{\bibfnamefont{E.}~\bibnamefont{Marinari}},
  \bibinfo{author}{\bibfnamefont{V.}~\bibnamefont{Martin-Mayor}},
  \bibnamefont{et~al.} (\bibinfo{collaboration}{Janus Collaboration}),
  \bibinfo{journal}{Phys. Rev. Lett.} \textbf{\bibinfo{volume}{120}},
  \bibinfo{pages}{267203} (\bibinfo{year}{2018}),
  \urlprefix\url{https://link.aps.org/doi/10.1103/PhysRevLett.120.267203}.

\bibitem[{\citenamefont{Amit and Amit}(1992)}]{amit1992modeling}
\bibinfo{author}{\bibfnamefont{D.~J.} \bibnamefont{Amit}} \bibnamefont{and}
  \bibinfo{author}{\bibfnamefont{D.~J.} \bibnamefont{Amit}},
  \emph{\bibinfo{title}{Modeling brain function: The world of attractor neural
  networks}} (\bibinfo{publisher}{Cambridge university press},
  \bibinfo{year}{1992}).

\bibitem[{\citenamefont{Ghofraniha et~al.}(2015)\citenamefont{Ghofraniha,
  Viola, Di~Maria, Barbarella, Gigli, Leuzzi, and
  Conti}}]{ghofraniha2015experimental}
\bibinfo{author}{\bibfnamefont{N.}~\bibnamefont{Ghofraniha}},
  \bibinfo{author}{\bibfnamefont{I.}~\bibnamefont{Viola}},
  \bibinfo{author}{\bibfnamefont{F.}~\bibnamefont{Di~Maria}},
  \bibinfo{author}{\bibfnamefont{G.}~\bibnamefont{Barbarella}},
  \bibinfo{author}{\bibfnamefont{G.}~\bibnamefont{Gigli}},
  \bibinfo{author}{\bibfnamefont{L.}~\bibnamefont{Leuzzi}}, \bibnamefont{and}
  \bibinfo{author}{\bibfnamefont{C.}~\bibnamefont{Conti}},
  \bibinfo{journal}{Nature communications} \textbf{\bibinfo{volume}{6}},
  \bibinfo{pages}{6058} (\bibinfo{year}{2015}).

\bibitem[{\citenamefont{Antenucci}(2016)}]{antenucci2016}
\bibinfo{author}{\bibfnamefont{F.}~\bibnamefont{Antenucci}},
  \emph{\bibinfo{title}{{\it {Statistical physics of wave interactions}}}}
  (\bibinfo{publisher}{Springer}, \bibinfo{year}{2016}).

\bibitem[{\citenamefont{Halasz et~al.}(1998)\citenamefont{Halasz, Jackson,
  Shrock, Stephanov, and Verbaarschot}}]{halasz1998phase}
\bibinfo{author}{\bibfnamefont{M.}~\bibnamefont{Halasz}},
  \bibinfo{author}{\bibfnamefont{A.}~\bibnamefont{Jackson}},
  \bibinfo{author}{\bibfnamefont{R.}~\bibnamefont{Shrock}},
  \bibinfo{author}{\bibfnamefont{M.~A.} \bibnamefont{Stephanov}},
  \bibnamefont{and}
  \bibinfo{author}{\bibfnamefont{J.}~\bibnamefont{Verbaarschot}},
  \bibinfo{journal}{Physical Review D} \textbf{\bibinfo{volume}{58}},
  \bibinfo{pages}{096007} (\bibinfo{year}{1998}).

\bibitem[{\citenamefont{Erik}(Oct. 2018)}]{ErikHormannT}
\bibinfo{author}{\bibfnamefont{H.}~\bibnamefont{Erik}},
  \emph{\bibinfo{title}{Optical spin glasses : a new model for glassy systems}}
  (\bibinfo{publisher}{Master Thesis, Univ. Sapienza}, \bibinfo{year}{Oct.
  2018}).

\bibitem[{\citenamefont{Pierangeli et~al.}(2020)\citenamefont{Pierangeli,
  Rafayelyan, Conti, and Gigan}}]{pierangeli2020scalable}
\bibinfo{author}{\bibfnamefont{D.}~\bibnamefont{Pierangeli}},
  \bibinfo{author}{\bibfnamefont{M.}~\bibnamefont{Rafayelyan}},
  \bibinfo{author}{\bibfnamefont{C.}~\bibnamefont{Conti}}, \bibnamefont{and}
  \bibinfo{author}{\bibfnamefont{S.}~\bibnamefont{Gigan}},
  \bibinfo{journal}{arXiv preprint arXiv:2006.00828}  (\bibinfo{year}{2020}).

\bibitem[{\citenamefont{Hopfield}(1982)}]{hopfield1982}
\bibinfo{author}{\bibfnamefont{J.~J.} \bibnamefont{Hopfield}},
  \bibinfo{journal}{Proceedings of the National Academy of Sciences}
  \textbf{\bibinfo{volume}{79}}, \bibinfo{pages}{2554} (\bibinfo{year}{1982}),
  ISSN \bibinfo{issn}{0027-8424}.

\bibitem[{\citenamefont{Hopfield}(1984)}]{hopfield1984}
\bibinfo{author}{\bibfnamefont{J.~J.} \bibnamefont{Hopfield}},
  \bibinfo{journal}{Proceedings of the National Academy of Sciences}
  \textbf{\bibinfo{volume}{81}}, \bibinfo{pages}{3088} (\bibinfo{year}{1984}),
  ISSN \bibinfo{issn}{0027-8424}.

\bibitem[{\citenamefont{Amit et~al.}(1985{\natexlab{a}})\citenamefont{Amit,
  Gutfreund, and Sompolinsky}}]{amit1985storing}
\bibinfo{author}{\bibfnamefont{D.~J.} \bibnamefont{Amit}},
  \bibinfo{author}{\bibfnamefont{H.}~\bibnamefont{Gutfreund}},
  \bibnamefont{and}
  \bibinfo{author}{\bibfnamefont{H.}~\bibnamefont{Sompolinsky}},
  \bibinfo{journal}{Physical Review Letters} \textbf{\bibinfo{volume}{55}},
  \bibinfo{pages}{1530} (\bibinfo{year}{1985}{\natexlab{a}}).

\bibitem[{\citenamefont{Amit et~al.}(1985{\natexlab{b}})\citenamefont{Amit,
  Gutfreund, and Sompolinsky}}]{amit1985spin}
\bibinfo{author}{\bibfnamefont{D.}~\bibnamefont{Amit}},
  \bibinfo{author}{\bibfnamefont{H.}~\bibnamefont{Gutfreund}},
  \bibnamefont{and}
  \bibinfo{author}{\bibfnamefont{H.}~\bibnamefont{Sompolinsky}},
  \bibinfo{journal}{Phys. Rev. A} \textbf{\bibinfo{volume}{32}},
  \bibinfo{pages}{1007} (\bibinfo{year}{1985}{\natexlab{b}}).

\bibitem[{\citenamefont{Amit et~al.}(1987)\citenamefont{Amit, Gutfreund, and
  Sompolinsky}}]{amit1987statistical}
\bibinfo{author}{\bibfnamefont{D.~J.} \bibnamefont{Amit}},
  \bibinfo{author}{\bibfnamefont{H.}~\bibnamefont{Gutfreund}},
  \bibnamefont{and}
  \bibinfo{author}{\bibfnamefont{H.}~\bibnamefont{Sompolinsky}},
  \bibinfo{journal}{Annals of Physics} \textbf{\bibinfo{volume}{173}},
  \bibinfo{pages}{30} (\bibinfo{year}{1987}).

\bibitem[{\citenamefont{Metropolis et~al.}(1953)\citenamefont{Metropolis,
  Rosenbluth, Rosenbluth, Teller, and Teller}}]{metropolis1953equation}
\bibinfo{author}{\bibfnamefont{N.}~\bibnamefont{Metropolis}},
  \bibinfo{author}{\bibfnamefont{A.~W.} \bibnamefont{Rosenbluth}},
  \bibinfo{author}{\bibfnamefont{M.~N.} \bibnamefont{Rosenbluth}},
  \bibinfo{author}{\bibfnamefont{A.~H.} \bibnamefont{Teller}},
  \bibnamefont{and} \bibinfo{author}{\bibfnamefont{E.}~\bibnamefont{Teller}},
  \bibinfo{journal}{The journal of chemical physics}
  \textbf{\bibinfo{volume}{21}}, \bibinfo{pages}{1087} (\bibinfo{year}{1953}).

\bibitem[{\citenamefont{Crisanti and Leuzzi}(2007)}]{crisanti2007equilibrium}
\bibinfo{author}{\bibfnamefont{A.}~\bibnamefont{Crisanti}} \bibnamefont{and}
  \bibinfo{author}{\bibfnamefont{L.}~\bibnamefont{Leuzzi}},
  \bibinfo{journal}{Phys. Rev. B} \textbf{\bibinfo{volume}{75}},
  \bibinfo{pages}{144301} (\bibinfo{year}{2007}).

\bibitem[{\citenamefont{G{\"o}tze}(2009)}]{goetze2009complex}
\bibinfo{author}{\bibfnamefont{W.}~\bibnamefont{G{\"o}tze}},
  \emph{\bibinfo{title}{Complex Dynamics of Glass-Forming Liquids, A
  Mode-Couplig Theory}} (\bibinfo{publisher}{Oxford University Press},
  \bibinfo{year}{2009}).

\bibitem[{\citenamefont{Caltagirone
  et~al.}(2012{\natexlab{a}})\citenamefont{Caltagirone, Ferrari, Leuzzi,
  Parisi, Ricci-Tersenghi, and Rizzo}}]{caltagirone2012critical}
\bibinfo{author}{\bibfnamefont{F.}~\bibnamefont{Caltagirone}},
  \bibinfo{author}{\bibfnamefont{U.}~\bibnamefont{Ferrari}},
  \bibinfo{author}{\bibfnamefont{L.}~\bibnamefont{Leuzzi}},
  \bibinfo{author}{\bibfnamefont{G.}~\bibnamefont{Parisi}},
  \bibinfo{author}{\bibfnamefont{F.}~\bibnamefont{Ricci-Tersenghi}},
  \bibnamefont{and} \bibinfo{author}{\bibfnamefont{T.}~\bibnamefont{Rizzo}},
  \bibinfo{journal}{Phys. Rev. Lett.} \textbf{\bibinfo{volume}{108}},
  \bibinfo{pages}{085702} (\bibinfo{year}{2012}{\natexlab{a}}).

\bibitem[{\citenamefont{Caltagirone
  et~al.}(2012{\natexlab{b}})\citenamefont{Caltagirone, Ferrari, Leuzzi,
  Parisi, and Rizzo}}]{caltagirone2012criticalslo}
\bibinfo{author}{\bibfnamefont{F.}~\bibnamefont{Caltagirone}},
  \bibinfo{author}{\bibfnamefont{U.}~\bibnamefont{Ferrari}},
  \bibinfo{author}{\bibfnamefont{L.}~\bibnamefont{Leuzzi}},
  \bibinfo{author}{\bibfnamefont{G.}~\bibnamefont{Parisi}}, \bibnamefont{and}
  \bibinfo{author}{\bibfnamefont{T.}~\bibnamefont{Rizzo}},
  \bibinfo{journal}{Phys. Rev. B} \textbf{\bibinfo{volume}{86}},
  \bibinfo{pages}{064204} (\bibinfo{year}{2012}{\natexlab{b}}).

\bibitem[{\citenamefont{Ferrari et~al.}(2012)\citenamefont{Ferrari, Leuzzi,
  Parisi, and Rizzo}}]{caltagirone2012twostep}
\bibinfo{author}{\bibfnamefont{U.}~\bibnamefont{Ferrari}},
  \bibinfo{author}{\bibfnamefont{L.}~\bibnamefont{Leuzzi}},
  \bibinfo{author}{\bibfnamefont{G.}~\bibnamefont{Parisi}}, \bibnamefont{and}
  \bibinfo{author}{\bibfnamefont{T.}~\bibnamefont{Rizzo}},
  \bibinfo{journal}{Phys. Rev. B} \textbf{\bibinfo{volume}{86}},
  \bibinfo{pages}{014204} (\bibinfo{year}{2012}).

\bibitem[{\citenamefont{Popoff et~al.}(2010{\natexlab{a}})\citenamefont{Popoff,
  Lerosey, Carminati, Fink, Boccara, and Gigan}}]{popoff2010measuring}
\bibinfo{author}{\bibfnamefont{S.}~\bibnamefont{Popoff}},
  \bibinfo{author}{\bibfnamefont{G.}~\bibnamefont{Lerosey}},
  \bibinfo{author}{\bibfnamefont{R.}~\bibnamefont{Carminati}},
  \bibinfo{author}{\bibfnamefont{M.}~\bibnamefont{Fink}},
  \bibinfo{author}{\bibfnamefont{A.}~\bibnamefont{Boccara}}, \bibnamefont{and}
  \bibinfo{author}{\bibfnamefont{S.}~\bibnamefont{Gigan}},
  \bibinfo{journal}{Physical review letters} \textbf{\bibinfo{volume}{104}},
  \bibinfo{pages}{100601} (\bibinfo{year}{2010}{\natexlab{a}}).

\bibitem[{\citenamefont{Hebb}(1962)}]{hebb1962organization}
\bibinfo{author}{\bibfnamefont{D.~O.} \bibnamefont{Hebb}},
  \emph{\bibinfo{title}{The organization of behavior: a neuropsychological
  theory}} (\bibinfo{publisher}{Science Editions}, \bibinfo{year}{1962}).

\bibitem[{\citenamefont{Vellekoop and Mosk}(2007)}]{vellekoop2007focusing}
\bibinfo{author}{\bibfnamefont{I.~M.} \bibnamefont{Vellekoop}}
  \bibnamefont{and} \bibinfo{author}{\bibfnamefont{A.}~\bibnamefont{Mosk}},
  \bibinfo{journal}{Optics letters} \textbf{\bibinfo{volume}{32}},
  \bibinfo{pages}{2309} (\bibinfo{year}{2007}).

\bibitem[{\citenamefont{Popoff et~al.}(2010{\natexlab{b}})\citenamefont{Popoff,
  Lerosey, Carminati, Fink, Boccara, and Gigan}}]{PhysRevLett.104.100601}
\bibinfo{author}{\bibfnamefont{S.~M.} \bibnamefont{Popoff}},
  \bibinfo{author}{\bibfnamefont{G.}~\bibnamefont{Lerosey}},
  \bibinfo{author}{\bibfnamefont{R.}~\bibnamefont{Carminati}},
  \bibinfo{author}{\bibfnamefont{M.}~\bibnamefont{Fink}},
  \bibinfo{author}{\bibfnamefont{A.~C.} \bibnamefont{Boccara}},
  \bibnamefont{and} \bibinfo{author}{\bibfnamefont{S.}~\bibnamefont{Gigan}},
  \bibinfo{journal}{Phys. Rev. Lett.} \textbf{\bibinfo{volume}{104}},
  \bibinfo{pages}{100601} (\bibinfo{year}{2010}{\natexlab{b}}).

\bibitem[{\citenamefont{Mattis}(1976)}]{mattis1976}
\bibinfo{author}{\bibfnamefont{D.}~\bibnamefont{Mattis}},
  \bibinfo{journal}{Phys. Lett.} \textbf{\bibinfo{volume}{56A}},
  \bibinfo{pages}{421} (\bibinfo{year}{1976}).

\bibitem[{\citenamefont{Buhmann and Schulten}(1987)}]{buhmann1987influence}
\bibinfo{author}{\bibfnamefont{J.}~\bibnamefont{Buhmann}} \bibnamefont{and}
  \bibinfo{author}{\bibfnamefont{K.}~\bibnamefont{Schulten}},
  \bibinfo{journal}{Biological cybernetics} \textbf{\bibinfo{volume}{56}},
  \bibinfo{pages}{313} (\bibinfo{year}{1987}).

\bibitem[{\citenamefont{Kob et~al.}(1997)\citenamefont{Kob, Donati, Plimpton,
  Poole, and Glotzer}}]{kob1997dynamical}
\bibinfo{author}{\bibfnamefont{W.}~\bibnamefont{Kob}},
  \bibinfo{author}{\bibfnamefont{C.}~\bibnamefont{Donati}},
  \bibinfo{author}{\bibfnamefont{S.~J.} \bibnamefont{Plimpton}},
  \bibinfo{author}{\bibfnamefont{P.~H.} \bibnamefont{Poole}}, \bibnamefont{and}
  \bibinfo{author}{\bibfnamefont{S.~C.} \bibnamefont{Glotzer}},
  \bibinfo{journal}{Physical review letters} \textbf{\bibinfo{volume}{79}},
  \bibinfo{pages}{2827} (\bibinfo{year}{1997}).

\bibitem[{\citenamefont{Beenakker}(1997)}]{beenakker1997random}
\bibinfo{author}{\bibfnamefont{C.~W.} \bibnamefont{Beenakker}},
  \bibinfo{journal}{Reviews of modern physics} \textbf{\bibinfo{volume}{69}},
  \bibinfo{pages}{731} (\bibinfo{year}{1997}).

\end{thebibliography}
\end{document}